\documentclass{aa}
\usepackage[varg]{txfonts}
\usepackage{braket}	
\usepackage{amssymb}
\usepackage[T1]{fontenc}
\usepackage{wrapfig}
\usepackage{fixltx2e}
\usepackage{mathtools}
\usepackage{array}
\usepackage{esint}
\usepackage{booktabs}
\usepackage{color}
\usepackage{enumitem}

\begin{document}

\title{Magnetic field transport in compact binaries}
 \author{N. Scepi\inst{1,2},  G. Lesur\inst{1}, G. Dubus\inst{1} and J. Jacquemin-Ide\inst{1}}
 \authorrunning{N. Scepi et al.}
   \institute{Univ. Grenoble Alpes, CNRS, Institut de Plan\'etologie et d'Astrophysique de Grenoble (IPAG), F-38000, Grenoble, France \and JILA, University of Colorado and National Institute of Standards and Technology, 440 UCB, Boulder, CO 80309-0440, USA
 \email{nisc6580@colorado.edu}
}

\abstract
{Dwarf nov\ae\ (DNe) and low mass X-ray binaries (LMXBs) show eruptions that are thought to be due to a thermal-viscous instability in their accretion disk. These eruptions provide constraints on angular momentum transport mechanisms.}
{We explore the idea that angular momentum transport could be controlled by the dynamical evolution of the large scale magnetic field. We study the impact of different prescriptions for the magnetic field evolution on the dynamics of the disk. This is a first step in confronting the theory of magnetic field transport with observations.}
{We develop a version of the disk instability model that evolves the density, the temperature and the large scale vertical magnetic flux together. We take into account the accretion driven by turbulence or by a magnetized outflow with prescriptions taken respectively from shearing box simulations or self-similar solutions of magnetized outflows. To evolve the magnetic flux, we use a toy model with physically motivated prescriptions depending mainly on the local magnetization $\beta$, where $\beta$ is the ratio of thermal pressure to magnetic pressure.}
{We find that allowing magnetic flux to be advected inwards provides the best agreement with DNe lightcurves. This leads to a hybrid configuration with an inner magnetized disk, driven by angular momentum losses to an MHD outflow, sharply transiting to an outer weakly-magnetized turbulent disk, where the eruptions are triggered. The dynamical impact is equivalent to truncating a viscous disk so that it does not extend down to the compact object, with the truncation radius dependent on the magnetic flux and evolving as $\dot{M}^{-2/3}$.}
{Models of DNe and LMXB lightcurves typically require the outer, viscous disk to be truncated in order to match observations. There is no generic explanation for this truncation. 
We propose that it is a natural outcome of the presence of large-scale magnetic fields in both DNe and LMXBs, the magnetic flux accumulating towards the center to produce a magnetized disk with a fast accretion timescale.}

\maketitle

\section{Introduction}
Magnetic fields in accretion disks are largely unconstrained observationally. Yet, the magnetic configuration has tremendous importance to understand the observed properties and evolution of astrophysical disks. It has long been known that active galactic nuclei (AGN; \citealt{lister2009}, \citealt{blandford2019}), protoplanetary disks (\citealt{burrows1996}, \citealt{ray1996}, \citealt{hirth1997},  \citealt{dougados2000}) and low-mass X-ray binaries (LMXBs; \citealt{mirabel1992}, \citealt{fender1997}) exhibit the presence of strongly collimated, fast outflows, called jets. It has recently been proposed that jets may also be present in dwarf nov\ae\ (DNe; \citealt{kording2008}, \citealt{russell2016}, \citealt{coppejans2016}, \citealt{coppejans2020}). These jets could be launched thanks to a strong (near equipartition) large scale poloidal magnetic field threading the disk that magneto-centrifugally accelerates and collimates the flow (\citealt{blandford1982}, \citealt{pudritz1983}, \citealt{konigl1989}, \citealt{ferreira1995}). More extended, slower ejections, called winds, have also been observed in all the accretion disks cited above (\citealt{cordova1982}, \citealt{mauche1987}, \citealt{crenshaw2003}, \citealt{miller2004}, \citealt{ponti2012}, \citealt{louvet2018}).

 Recently, 3D MRI numerical simulations (\citealt{fromang2013}, \citealt{bai2013b}, \citealt{lesur2013}) and self-similar solutions \citep{jacquemin2019} showed that outflows \textit{\`a la} \cite{blandford1982} could be extended to lower magnetization to produce slower, more massive solutions compared to the previously computed jet solutions \citep{ferreira1997}. However, it is not clear that these lower magnetization outflows could entirely explain the observed features of winds. Indeed, at least in DNe and LMXBs, it seems that to reproduce the estimated mass loss rates and terminal velocities from observations there should be some hybrid mechanism involving magnetic and thermal driving or radiative driving (\citealt{proga2003}, \citealt{miller2006}, \citealt{luketic2010}, \citealt{higginbottom2015}, \citealt{chakravorty2016}, \citealt{diaz2016}, \citealt{jacquemin2019},\citealt{trueba2019}). Nonetheless, both jets and winds in accretion disks depend strongly on the presence of a large scale magnetic field threading the disk.

A large scale magnetic field does not only impact the ejection mechanisms but also the accretion processes. In fact, the magnetic field is believed to be one of the main drivers of accretion, either through turbulence due to the magneto-rotational instability (MRI; \citealt{velikhov1959}, \citealt{chandrasekhar1960}, \citealt{balbus1991})\footnote{We note that MRI does not require a large scale magnetic field.}, which transports angular momentum radially, or through magnetic winds/jets (\citealt{blandford1982}, \citealt{pudritz1983}, \citealt{konigl1989}, \citealt{ferreira1995}), which extract angular momentum vertically. Moreover, both the strength of MRI (\citealt{hawley1995}, \citealt{salvessen2016}, \citealt{scepi2018b}) and the properties of the magnetic outflows \citep{jacquemin2019} depend on the large scale poloidal magnetic field strength.\\

Large scale poloidal magnetic fields can have two origins. They could either be generated in situ by dynamo effects (\citealt{brandenburg1995}, \citealt{tout1996}) or could be advected inwards from the surrounding medium \citep{bisnovatyi1974}. It is still not clear whether MRI dynamo can produce an organized poloidal magnetic field on large scales (\citealt{brandenburg1995}). Local shearing box simulations tend to show that they do not but state-of-the-art 3D global simulations from \cite{liska2018} found that loops of poloidal magnetic field when formed in the outer region and dragged in the inner region of the disk becomes "large scale" due to the difference of length scale between the outer and inner region.

In a seminal work, \cite{lubow1994} studied the transport of magnetic field in a thin disk where angular momentum transport is only due to turbulence, with turbulence prescribed as an $\alpha$ parameter, where $\alpha$ is the ratio of the radial stress to the thermal pressure \citep{Shakura}. They showed that in an $\alpha$-disk, it seems impossible to advect large scale magnetic field inwards. These results have been later confirmed by \cite{heyvaerts1996}. Indeed, with a magnetic Prandtl number (ratio of the turbulent viscosity to the magnetic turbulent diffusivity) of unity, as seems to be the case for MRI-driven turbulence \citep{lesur2009}, the  large scale poloidal magnetic field diffuses outwards efficiently. \cite{lubow1994} considered the advection of magnetic field inwards by turbulent accretion and computed the diffusion term by solving the global magnetic structure in the potential field approximation. However, the authors ignored the vertical structure of the disk and assumed that the velocity at which the magnetic flux is advected is the same as the velocity at which the mass is accreted. 

\cite{bisnovatyi2007} and \cite{rothstein2008} proposed that taking into account the vertical structure of the disk could help advect magnetic field inwards. Their idea is based on the fact that the upper layers of the disk, where the MRI is suppressed, have a higher effective conductivity than the bulk of the disk. Although, accretion in the upper layers does not participate much to mass transport, because the density is low in those regions, it could help advect magnetic field inwards faster than it would do in the more diffusive midplane of the disk. Evidence for this mechanism is also found in 3D global simulations (\citealt{beckwith2009}, \citealt{tchekhovskoy2011}, \citealt{suzuki2014}, \citealt{zhu2018}) where funnels of low-density matter above the disk carry the magnetic flux inwards. \cite{guilet2012} computed the vertical structure of a thin disk through an asymptotic expansion of the resistive MHD equations. With this semi-local model, they were able to compute the transport rates due to an $\alpha$ turbulent accretion as well as MHD outflow-driven accretion as a function of magnetization. They showed, in the context of protoplanetary disks, that by taking account the vertical structure of the disk the magnetic field could be advected efficiently inward for a realistic magnetic Prandtl number because of accretion in the upper low-density layers \citep{guilet2014}. 

The problem of inefficient dragging of the magnetic field is fundamentally due to the fact that the  turbulent magnetic diffusivity is of the same order than the turbulent viscosity \citep{tout1996}. This problem could be alleviated if the angular momentum is removed by the magnetic outflow-driven torque, which can be more efficient than the turbulent torque (\citealt{lubow1994b}, \citealt{heyvaerts1996}, \citealt{ogilvie2001}, \citealt{guilet2012}). \cite{li2019} constructed a global model of thin disks where the removal of angular momentum is done by a magnetic outflow and the magnetic structure is evolved self-consistently. To evolve the magnetic field, they computed the global magnetic structure in the same way as \cite{lubow1994}. They showed that magnetic field is advected inwards very efficiently and forms a very magnetised zone around the central object even when they start from a relatively moderate external magnetic field. Their work focused on the stationary, global structure of the disk and the outflows. 

In this paper, we wish to explore the dynamical evolution of a disk threaded everywhere by a large scale poloidal magnetic field in the context of transient compact objects. We will focus, in particular, on the case of dwarf nov\ae, which are binary systems composed of a white dwarf and a solar type star, where the accretion disk surrounding the white dwarf undergoes thermal-viscous eruptions. The eruptions of dwarf nov\ae\ have been modeled for years using the disk instability model (DIM, \citealt{lasota2001}). This model traditionally assumes that the transport of angular momentum is due to turbulence. Dwarf nov\ae\ have been used as a testbed to deduce the value of $\alpha$ from observations (\citealt{kotko2012}, \citealt{cannizzo2012}). However, despite many efforts, no physically satisfying solution was found to produce the values of $\alpha$ required by observations (\citealt{hirose2014}, \citealt{coleman2016}, \citealt{scepi2018}). Recently, \cite{scepi2019} proposed that the torque produced by a magnetic outflow could help solve this issue. They showed that the inclusion of the outflow-driven magnetic torque shapes the light curves. In particular, they were able to produce realistic eruptions by assuming the magnetic field $B$ was a function of disk radius $B\propto R^{-3}$.
However, \cite{scepi2019} fixed the magnetic configuration, whereas it is expected to evolve during the outburst as the magnetic field is advected and diffuses in the disk. We wish here to take this into account by evolving self-consistently the large scale magnetic configuration together with the modified disk instability model of \cite{scepi2019}. We emphasize that we do not aim at comparing directly our results with observations of DNe but rather to provide a first theoretical study of the interplay between magnetic field transport and the eruptions of DNe. We also stress that we are not  modelling the magnetospheric interaction of a possible white dwarf dipolar field with the disk. This is usually captured as a truncation of the inner disk at the magnetospheric radius, the radius below which the dipolar field is strong enough to funnel accretion \citep{Frank}. Such a truncation has been proposed, in combination with the DIM, to explain the high X-ray flux in quiescence and the lag between the outburst rise in UV and optical in DNe \citep{1992MNRAS.259P..23L,2003A&A...410..239S}.

In \S\ref{sec:methods} we expose in detail the method we use to evolve the fields in our model. In \S\ref{sec:efficient} and \S\ref{sec:noadvection}, we explore the behavior of the disk using two different prescriptions for the magnetic field evolution. We show the impact of this choice on the light curves of DNe. Then, in \S\ref{sec:discussion} we discuss the caveats of our model, its observational consequences and its application to LMXBs. We conclude in \S\ref{sec:conclusion}.

\section{Methods}\label{sec:methods}
In this paper, we consider an axisymetric accretion disk around a white dwarf of 0.6 $M_{\odot}$. To study the evolution of the ionized plasma that constitutes the accretion disk we use the viscous and resistive non-relativistic MHD equations in a frame centered on the white dwarf and use cylindrical coordinates. 
\subsection{Induction equation}
In a axisymetric disk, we can write the magnetic field $\textbf{B}$ as 
\begin{equation}\label{eq:B_psi}
\textbf{B}=-\frac{1}{R}\textbf{e}_\phi\times\pmb{\nabla}\psi+B_\phi\textbf{e}_\phi
\end{equation}
where $\psi$ is the magnetic flux function and $R$ the cylindrical radius in the disk. The induction equation is written as 
\begin{equation}\label{eq:induction}
\frac{\partial\textbf{B}}{\partial t}=\nabla\times(\textbf{v}\times\textbf{B}-\eta\nabla\times\textbf{B})
\end{equation}
where $\textbf{v}$ is the velocity and $\eta$ is the magnetic diffusivity, here assumed to be of turbulent origin.

We can rewrite (\ref{eq:induction}) using (\ref{eq:B_psi}) and find 
\begin{equation}\label{eq:dt_psi_i}
\frac{\partial \psi}{\partial t}=R\left[v_zB_R-v_RB_z-\eta\left(\frac{\partial B_R}{\partial z}-\frac{\partial B_z}{\partial R}\right)\right]=0.
\end{equation}

Following the asymptotic expansion in a thin disk of the viscous and resistive MHD equations developed by \cite{guilet2012}, we write
\begin{equation}\label{eq:psi_dec}
\psi(R,z,t)=\psi_0(R,t)+\epsilon^2\psi_2(R,z,t)+o(\epsilon^2)
\end{equation}
where $\epsilon$ is a small dimensionless parameter of the order of $H/R$. $H\equiv c_s/\Omega$ is the scale height of the disk, $c_s$ the sound speed and $\Omega$ the angular velocity. According to \cite{guilet2012}, we assume
\begin{equation}\label{eq:asymptotic}
\frac{v_z}{v_R}=O(\epsilon)\:;\:\:\:\:\frac{B_R}{B_z}=O(\epsilon).
\end{equation}
One can check that indeed
\begin{equation*}
\frac{B_R}{B_z}=\frac{-\frac{1}{R}\partial_z({\epsilon^2\psi_2})}{\frac{1}{R}\partial_R\psi_0}=O(\epsilon)
\end{equation*}
since $z/R=O(\epsilon)$ in the disk. We discuss the limitations of these assumptions in \S\ref{sec:limitations}.\\

Using (\ref{eq:B_psi}), (\ref{eq:dt_psi_i}), (\ref{eq:psi_dec}) and (\ref{eq:asymptotic}) we can rewrite the poloidal part of the induction equation at leading order as 
\begin{equation}\label{eq:psi_evol}
\frac{\partial \psi_0}{\partial t}+v_R\frac{\partial\psi_0}{\partial R}+\eta R\left(\frac{\partial B_R}{\partial z}-\frac{\partial B_z}{\partial R}\right)=0.
\end{equation}

We then integrate vertically equation (\ref{eq:psi_evol}) weighted by the conductivity $1/\eta$ \citep{ogilvie2001} to find
\begin{equation}\label{eq:psi_evol2}
\frac{\partial \psi_0}{\partial t}+ RB_z\left[\overline{v}_R+\overline{\eta}\left(\frac{1}{L_z}\frac{B_{Rs}}{B_z}-\frac{\partial\ln B_z}{\partial R}\right)\right]=0
\end{equation}
with
\begin{equation*}
\frac{1}{\overline{\eta}}=\frac{1}{2L_z}\int\frac{1}{\eta}dz,
\end{equation*}
 \begin{equation*}
\overline{v}_R=\frac{\overline{\eta}}{2L_z}\int\frac{v_R}{\eta}dz
\end{equation*}
 and $L_z$ is the height up to which we integrate in the vertical direction. $B_{Rs}$ is the radial magnetic field at the disk surface. We assumed an odd symmetry for the radial magnetic field and integrated all quantities between $-L_z$ and $L_z$.
Equation (\ref{eq:psi_evol2}) can be rewritten as 
\begin{equation}\label{eq:psi_adv_diff}
\frac{\partial \psi}{\partial t}+v_\psi \frac{\partial \psi}{\partial R}=0
\end{equation}
where 
\begin{equation}\label{eq:def_vpsi}
v_\psi\equiv\overline{v}_R+\overline{\eta}\left(\frac{1}{L_z}\frac{B_{Rs}}{B_z}-\frac{\partial\ln B_z}{\partial R}\right)
\end{equation}
is the transport velocity of the magnetic flux. We dropped the subscript $0$ in the magnetic flux function.
From (\ref{eq:psi_adv_diff}) it seems that the induction equation has become a simple advection problem. However, one should keep in mind that the velocity $v_\psi$ depends on $\partial_{R}B_z=\partial_{R}\left(\frac{1}{R}\frac{\partial\psi}{\partial R}\right)$ and on $B_{Rs}=-\frac{1}{R}\partial_{z}\psi|_s$ so that (\ref{eq:psi_adv_diff}) describes a advection/diffusion problem.

\subsection{Prescriptions for $v_\psi$}\label{sec:vpsi_constraints}
In the absence of a prescription for $v_\psi$ derived from first principles, we assume that magnetic flux transport is mainly local and can be prescribed as a function of $\beta$, with 
\begin{equation*}
\beta\equiv\frac{8\pi P_\mathrm{mid}}{B_z^2},
\end{equation*} 
where $P_\mathrm{mid}$ is the mid-plane thermal pressure. We write $\beta(z)$ as $8\pi P(z)/B_z^2$, otherwise $\beta$ is always computed at the mid-plane. We discuss the limitations of our local approach in \S\ref{sec:limitations}. 

With this model, we aim to study the impact on the dynamics of the disk of different prescriptions for $v_\psi$ . We also tried to use the more sophisticated linear model of \cite{guilet2012} to obtain a prescription for $v_\psi$ but encountered issues with this approach (see Appendix \ref{sec:appendixA}).

We decompose $v_\psi$ as 
\begin{equation}\label{eq:vpsi_pres}
v_\psi=-v_\mathrm{in}+v_\mathrm{out}-v_{D_{B_z}}D_{B_z},
\end{equation}
where $v_\mathrm{in}$ represents the inwards transport velocity of magnetic flux as the flux is dragged inwards with the accretion flow. In our case, we expect accretion to be due to the turbulent torque or a surface torque due to a magnetized outflow. 

Inversely, $v_\mathrm{out}$ represents the outward transport velocity of magnetic flux. It is expected to be dominated by the vertical diffusion of the radial magnetic field. 

Finally, $v_{D_{B_z}}D_{B_z}$ represents the radial diffusion of the vertical magnetic field, with $D_{B_z}\equiv\frac{\partial\ln B_z}{\partial \ln R}$. 

For simplicity, we impose $v_\mathrm{in}$ and $v_\mathrm{out}$ as power laws of $\beta$ so as to explore a wide range of regime for magnetic flux transport. To reduce the degrees of liberty concerning the choice of $v_\mathrm{in}$ and $v_\mathrm{out}$ we give here some expected order of magnitudes for the three velocities. 

\subsubsection{Constraints on $v_{D_{B_z}}$}
Following our simple approach we set 
\begin{equation*}
v_{D_{B_z}}D_{B_z}\equiv\overline{\eta}\frac{\partial\ln B_z}{\partial R}. 
\end{equation*}
Assuming that $\eta(z)$ is dominated by the effective turbulent resistivity and is constant with height, this can be rewritten as 
\begin{equation*}
v_{D_{B_z}}=\frac{\alpha}{\mathcal{P}}\frac{H}{R}c_s 
\end{equation*}
where 
\begin{equation}\label{eq:alpha_prescription}
\alpha=\mathrm{min}(1\:,\:15\times\beta^{-0.56}+0.03)
\end{equation}
is taken from the 3D shearing box simulations with radiative transfer of \cite{scepi2018b}, adding a cap in the low $\beta$ limit to mimic the fact that MRI-driven turbulence is quenched when $\beta<1$. We also introduce the magnetic Prandtl number,
$\mathcal{P}$, the ratio of the effective turbulent viscosity to the effective turbulent resistivity, which is set to unity in agreement with \cite{lesur2009}. 

Our expression for $v_{D_{B_z}}$ is a crude estimation since it does not take into account the impact of the radial gradient of vertical magnetic field on $v_R$ and assumes a constant resistivity.

\subsubsection{Constraints on $v_\mathrm{out}$}\label{sec:est_vout}
From (\ref{eq:def_vpsi}) we can estimate 
\begin{equation}\label{eq:vout_estimate}
v_\mathrm{out}\sim\frac{\alpha}{\mathcal{P}}\frac{H}{L_z}\frac{B_{Rs}}{B_z}c_s.
\end{equation}
$L_z$ represents the surface of the disk and we can crudely assume that it is of the order of $\sim H$. Constraining the radial surface magnetic field using the disk-magnetized outflow self-similar solutions of \cite{jacquemin2019} in the regime $1<\beta<10^4$, we have
\begin{equation}\label{eq:qr}
\frac{B_{Rs}}{B_z}=0.62\times\beta^{0.4}.
\end{equation}
Using (\ref{eq:alpha_prescription}), (\ref{eq:vout_estimate}) and (\ref{eq:qr}) we see that 
\begin{equation*}
v_\mathrm{out}\sim c_s
\end{equation*} 
when $\beta\approx1$.

We note that (\ref{eq:qr}) is not in agreement with our assumption that $B_R/B_z=O(\epsilon)$. We discuss the implication of this in \S\ref{sec:limitations}.

\subsubsection{Constraints on $v_\mathrm{in}$}\label{sec:est_vin}
It is more difficult to obtain a simple order of estimate for $v_\mathrm{in}$ as we do not know \textit{a\:priori} the vertical structure of $v_R$. In the model of \cite{guilet2012} the transport velocity of magnetic flux inwards, $v_\mathrm{in}$, due to either the turbulent torque or the surface torque from a magnetized outflow, tends towards the mass transport velocity, $\tilde{v}_R$, in the highly magnetized case (see their Figure 8). Indeed, in their model the height where the surface accretion occurs tends towards the midplane as $\beta$ decreases and the magnetic flux transport velocity tends towards the mass transport velocity. In the highly magnetized regime ($\beta\approx1$), the mass transport velocity is dominated by the torque from the magnetized outflow \citep{ferreira1995} with
\begin{equation*}
\tilde{v}_R=\frac{4q}{\beta}c_s,
\end{equation*}
where $q\equiv B_{\phi s}/B_z$ is given by the self-similar solutions of \cite{jacquemin2019}
\begin{equation}\label{eq:q}
q=0.36\times\beta^{0.6}.
\end{equation}
We see that when $\beta\approx 1$, the mass transport velocity, $\tilde{v}_R$, approaches unity and so 
\begin{equation*}
v_\mathrm{in}\sim c_s.
\end{equation*} 

\subsubsection{Behavior at large $\beta$}\label{sec:behavior_largebeta}
We expect $v_\mathrm{out}$ to represent the vertical diffusion of the radial magnetic field, where the surface radial magnetic field is set by a magnetized outflow. We argue here that an MHD outflow cannot exist at low magnetization and that subsequently all magnetic flux transport associated with an MHD outflow should decrease with $\beta$.

The base of the outflow (or surface of the disk) can be roughly estimated as the height, $\zeta_B$, where $\beta(z)=1$ (\citealt{suzuki2009}, \citealt{fromang2013}, \citealt{zhu2018}). So, as $\beta$ increases $\zeta_B$ increases too. However, in a real disk, the base of the outflow, where angular momentum is removed from the disk, must originate from a resistive region where the matter is able to decouple from the magnetic field to be accreted \citep{ferreira1995}. In a realistic model, the effective turbulent diffusivity should decrease with height as it is set by the turbulent profile in the disk. Hence, in a real disk, there should exist a critical value of $\beta$ above which an outflow cannot be launched. This critical $\beta$ corresponds to the magnetization where $\zeta_B$ becomes larger than the vertical extent of the resistive region.

In our estimations, we assumed that the turbulent resistivity, $\eta$, does not depend on $z$ and so we do not capture this effect. However, the expected limited vertical extent of the resistive region implies that $v_\mathrm{out}$ must decrease with $\beta$ and be null for high $\beta$. 

The case of $v_\mathrm{in}$ is more complicated since we expect inward advection to be due to both outflow driven accretion and MRI turbulent driven accretion. In contrast, to outflow driven accretion, turbulent MRI driven accretion is not expected to fade out as $\beta$ goes to infinity because of the MRI dynamo \citep{scepi2018b}. So, if we consider that turbulent driven accretion efficiently advects magnetic flux inward as in \cite{guilet2014}, $v_\mathrm{in}$ does not have to be null at high $\beta$. However, if we assume that turbulent driven accretion does not efficiently advect magnetic flux inward as in \cite{lubow1994}, $v_\mathrm{in}$ does have to be null at high $\beta$. We explore both options in \S\ref{sec:efficient} and \S\ref{sec:noadvection} respectively.\\

To summarize we have three constraints to build our prescriptions of $v_\psi$. First, $v_\mathrm{in}$ and $v_\mathrm{out}$ should be of the order of $c_s$ when $\beta$ approaches unity. Second, $v_{D_{B_z}}$ is smaller than $v_\mathrm{in}$ and $v_\mathrm{out}$ by a factor $H/R$ when $\beta$ approaches unity. Third, $v_\mathrm{out}$ (and $v_\mathrm{in}$ if we only consider accretion due to a magnetized outflow) decreases when increasing $\beta$ and goes to zero as $\beta$ tends to infinity.

\subsubsection{A variety of prescriptions}
\begin{figure}[h!]
\includegraphics[width=90mm,height=120mm]{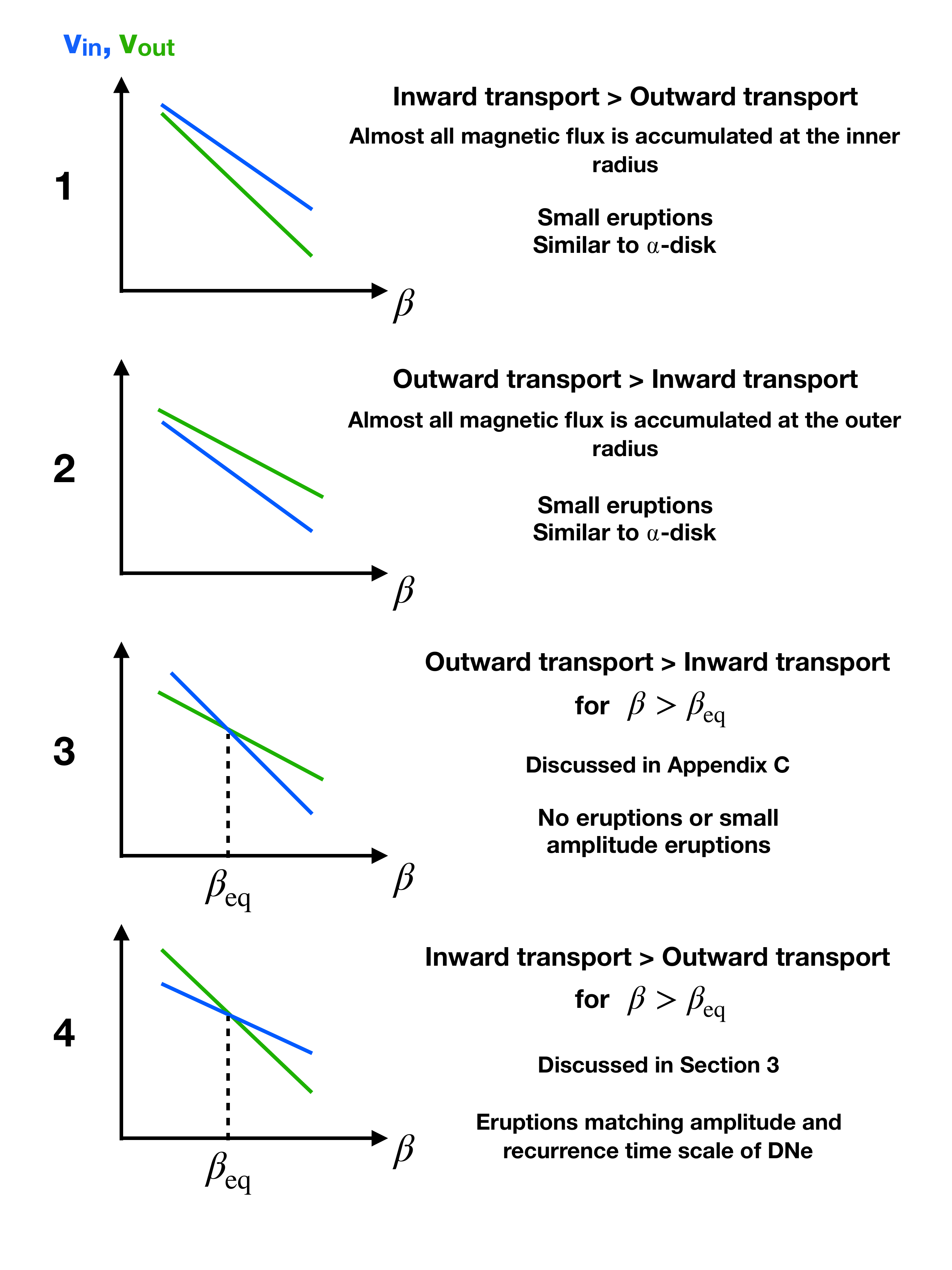}
\caption{Sketch of the four set of prescriptions obeying the constraints on $v_\mathrm{in}$ and $v_\mathrm{out}$ from \S\ref{sec:vpsi_constraints} with a comment on the resulting light curves and disk's configuration. Green lines represent outward transport and blue lines inward transport.}
\label{fig:vpsi_scenarii}
\end{figure}

When building prescriptions for $v_\mathrm{in}$ and $v_\mathrm{out}$ as a function of $\beta$, we can imagine two different scenarii : 
\begin{itemize}[label={--}]
\item  $v_\mathrm{in}$ and $v_\mathrm{out}$ are never equal and one dominates the other for every $\beta$ (cases 1 and 2 in Figure \ref{fig:vpsi_scenarii})
\item or $v_\mathrm{in}$ and $v_\mathrm{out}$ are equal at some $\beta$ that we call $\beta_\mathrm{eq}$ (cases 3 and 4 in Figure \ref{fig:vpsi_scenarii}).
\end{itemize} 

We find that the first two sub-cases are actually of low interest. Indeed in these cases, the magnetic field is accumulated either in the inner parts (case 1 in Figure \ref{fig:vpsi_scenarii}) or the outer parts (case 2 in Figure \ref{fig:vpsi_scenarii}) over a very small range in radii\footnote{Note that this is due to our assumption that $v_\psi$ is null at the boundaries to preserve the magnetic flux in the disk (see \S\ref{sec:DIM}). In the case where the magnetic flux would be able to leave the disk, case 1 and case 2 would lead to the disk emptying of magnetic flux from the inner boundary and outer boundary respectively.} . The range of radii over which the flux accumulates depends on the competition between $v_\mathrm{in}$ (or $v_\mathrm{out}$) and $v_{D_{B_z}}D_{B_z}$. Since $v_{D_{B_z}}$ is smaller than $v_\mathrm{in}$ (or $v_\mathrm{out}$) by a factor $\approx H/R$ in the magnetized regime, one needs to build a large radial gradient of $B_z$ before reaching a magnetic equilibrium. This explains that the field ends up concentrated within the innermost (outermost) radii only. The disk then becomes unmagnetized except for a highly magnetized inner (outer) zone. The highly magnetized zone does not evolve with time and the rest of the disk has small eruptions as expected in disks with constant $\alpha$.\\

The case where there exists an equilibrium magnetization, $\beta_\mathrm{eq}$, where the inner transport of magnetic flux is balanced by the outer transport of magnetic flux is of greater interest. One can imagine more complicated scenarii where $v_\mathrm{in}$ and $v_\mathrm{out}$ cross several times. We do not discuss these cases for simplicity.

Once we have chosen that there exists a $\beta_\mathrm{eq}$, one again faces two choices. 
\begin{itemize}[label={--}]
\item $v_\mathrm{out}>v_\mathrm{in}$ for $\beta>\beta_\mathrm{eq}$ (case 3 in Figure \ref{fig:vpsi_scenarii})
\item or $v_\mathrm{in}>v_\mathrm{out}$ for $\beta>\beta_\mathrm{eq}$ (case 4 in Figure \ref{fig:vpsi_scenarii}).
\end{itemize} 
As discussed in Appendix \ref{sec:appendixC}, we believe that case 4, where $v_\mathrm{in}>v_\mathrm{out}$ for $\beta>\beta_\mathrm{eq}$, is the astrophysically interesting case and we mainly discuss this case. A discussion on case 3 is added in Appendix \ref{sec:appendixC}.

\subsection{Modified DIM code with magnetic field transport}\label{sec:DIM}
To evolve the radial structure of the disk with time  we solve the vertically averaged equations of a thin disk :
\begin{equation}\label{eq:dt_sigma}
\frac{\partial \Sigma}{\partial t}=\frac{1}{2\pi R}\frac{\partial\dot{M}}{\partial R}
\end{equation}
\begin{equation}\label{eq:dt_Tc}
\frac{\partial T_c}{\partial t}=\frac{Q^+-Q^-}{\Sigma}\frac{2m_p}{5k_B\mu}
\end{equation}
and 
\begin{equation}\label{eq:dt_psi}
\frac{\partial \psi}{\partial t}=(v_\mathrm{in}-v_\mathrm{out}+v_{D_{B_z}}D_{B_z}) \frac{\partial \psi}{\partial R},
\end{equation}
where $\dot{M}$ is the mass accretion rate, $T_c$ the temperature in the midplane of the disk, $Q^+$ the heating rate, $Q^-$ the cooling rate, $m_p$ the mass of a proton, $k_B$ the Boltzman constant and $\mu$ the mean molecular weight. Note that for the sake of simplification, we have neglected mass losses induced by the wind in (\ref{eq:dt_sigma}).
We compute the mass accretion rate as
\begin{equation}
\dot{M}=\frac{4\pi}{R\Omega}\frac{\partial}{\partial R}\left(\frac{3}{2}R^2\alpha(\beta)\Sigma\Omega^2H^2\right)+q(\beta)B_z^2\frac{R}{\Omega}
\end{equation}
where we take into account the accretion due to MRI turbulence and the outflow magnetic torque on the surface of the disk. The value of $\alpha(\beta)$ is taken from (\ref{eq:alpha_prescription}). The value of $q$ is given by (\ref{eq:q}). The heating rates and cooling rates are given respectively by  
\begin{equation}\label{eq:heating}
Q^+=\frac{9}{4}\alpha\Omega^3H^2\Sigma
\end{equation} 
and  
\begin{equation}
Q^-=2\sigma T_\mathrm{eff}^4.
\end{equation}
The effective temperature $T_\mathrm{eff}$ is a function of $\Sigma$ and $\kappa$ and is computed using the prescriptions from \cite{latter2012}. We note that the treatment of the thermal structure in our model is simpler than in \cite{hameury1998} and \cite{scepi2019}. Indeed, it does not include cooling due to convection, the change in mean molecular weight and thermal capacity as hydrogen is ionizing and recombining, the radial transport of energy and the heating/cooling by adiabatic expansion of the fluid. In \cite{scepi2019} we found that the last two terms, in particular, were only important near sharp transitions in surface density or temperature and heating/cooling fronts (see also \citealt{menou1999}). Comparing our model with the one from \cite{scepi2019} we find that including these terms does not alter drastically the overall dynamics of the disk, although the eruptions tend to be longer, with a slower rise to the eruptive state, and exhibit reflares. We caution that the simplified cooling function of \cite{faulkner1983}, which inspired the prescription of \cite{latter2012} that we adopt here, is known to lead to outbursts with a larger amplitude than more realistic treatments (\citealt{lin1985}, \citealt{hameury1998}). This is especially true when the inner edge of the disk is large (see Figure 5 and 6 of \cite{lin1985} and \S\ref{sec:LC}). Nonetheless, we believe that our model captures most of the necessary physics to understand the coupling between magnetic field transport and the eruptions. We also note that (\ref{eq:heating}) overestimates the heating rate. Part of the gravitational energy that is usually extracted from the disk by the turbulent stress tensor is actually used to launch the outflow and goes in the vertical Poynting flux instead of being liberated into thermal energy locally \citep{ferreira1995}.\\

The novelty compared to \cite{scepi2019} is that we evolve the magnetic field at the same time as the surface density and the temperature by solving (\ref{eq:dt_psi}). Equation (\ref{eq:dt_psi}) is coupled to (\ref{eq:dt_sigma}) and (\ref{eq:dt_Tc}) through the prescriptions for $v_\mathrm{in}$, $v_\mathrm{out}$, and $v_\mathrm{D_B},$ which depends on $\beta$ and $H/R$. We also use a different prescription for the magnetic outflow's torque, $q$, taken from the self similar solutions of \cite{jacquemin2019}. While prescriptions from shearing box simulations concerning the turbulent torque are believed to be quite accurate the prescriptions concerning the magnetic outflow's torque might be better captured by self-similar solutions which take into account the global curvature of the outflow, which is ignored in local shearing box models. In any case, all the results presented here do not depend on the choice of $q(\beta)$.\\

We start from a uniform magnetic field configuration. The disk is isolated, meaning that it cannot lose or gain magnetic flux as we enforce $v_\psi$ to cancel at the inner and outer boundaries. We set up $\Sigma$ to be small at the inner boundary (equal to 1) and the mass accretion rate to be equal to the external mass accretion rate $\dot{M}_\mathrm{ext}=10^{16}\:\mathrm{g\:s^{-1}}$ at the outer boundary. The internal radius of the disk is $R_\mathrm{in}=8\times10^{8}\:\mathrm{cm}$, the external radius of the disk is fixed at $R_\mathrm{out}=2\times10^{10}\:\mathrm{cm}$ and the mass of the white dwarf is 0.6 $M_\odot$. We use a static logarithmic grid in radius with a resolution 200 points. We went up to a resolution of 800 points and did not see any meaningful difference concerning the overall dynamics of the disk.

\section{Magnetic advection dominates for $\beta>\beta_\mathrm{eq}$}
In this section, we study two cases:
\begin{itemize}[label={--}]
\item  turbulent accretion is efficient in advecting magnetic flux inwards at large $\beta$ (as in \citealt{guilet2014}) and so $v_\mathrm{in}>v_{D_{B_z}}$  at large $\beta$ (blue solid line in Figure \ref{fig:vpsi_pres} and \S\ref{sec:efficient}).
\item turbulent accretion is not efficient in advecting magnetic flux inwards at large $\beta$ (as in \citealt{lubow1994}) and so $v_\mathrm{in}<v_{D_{B_z}}$  at large $\beta$ (blue dashed line in Figure \ref{fig:vpsi_pres} and \S\ref{sec:noadvection}). 
\end{itemize}

\subsection{Efficient advection at large $\beta$}\label{sec:efficient}
We start with the case where turbulent accretion is efficient in advecting magnetic flux inwards at large $\beta$. For this subsection, we use the following prescriptions for $v_\mathrm{in}$ and $v_\mathrm{out}$:
\begin{gather}
v_\mathrm{in}=3\times\beta^{-0.2} c_s \nonumber \\ 
v_\mathrm{out}=4f\times\beta^{-0.3} c_s \label{eq:vout}
\end{gather}
where $f=1/(1+10^{-10}\beta^2)$ is used to quench the outflow at large $\beta$ (see \S\ref{sec:behavior_largebeta}).

\begin{figure}[h!]
\includegraphics[width=90mm,height=70mm]{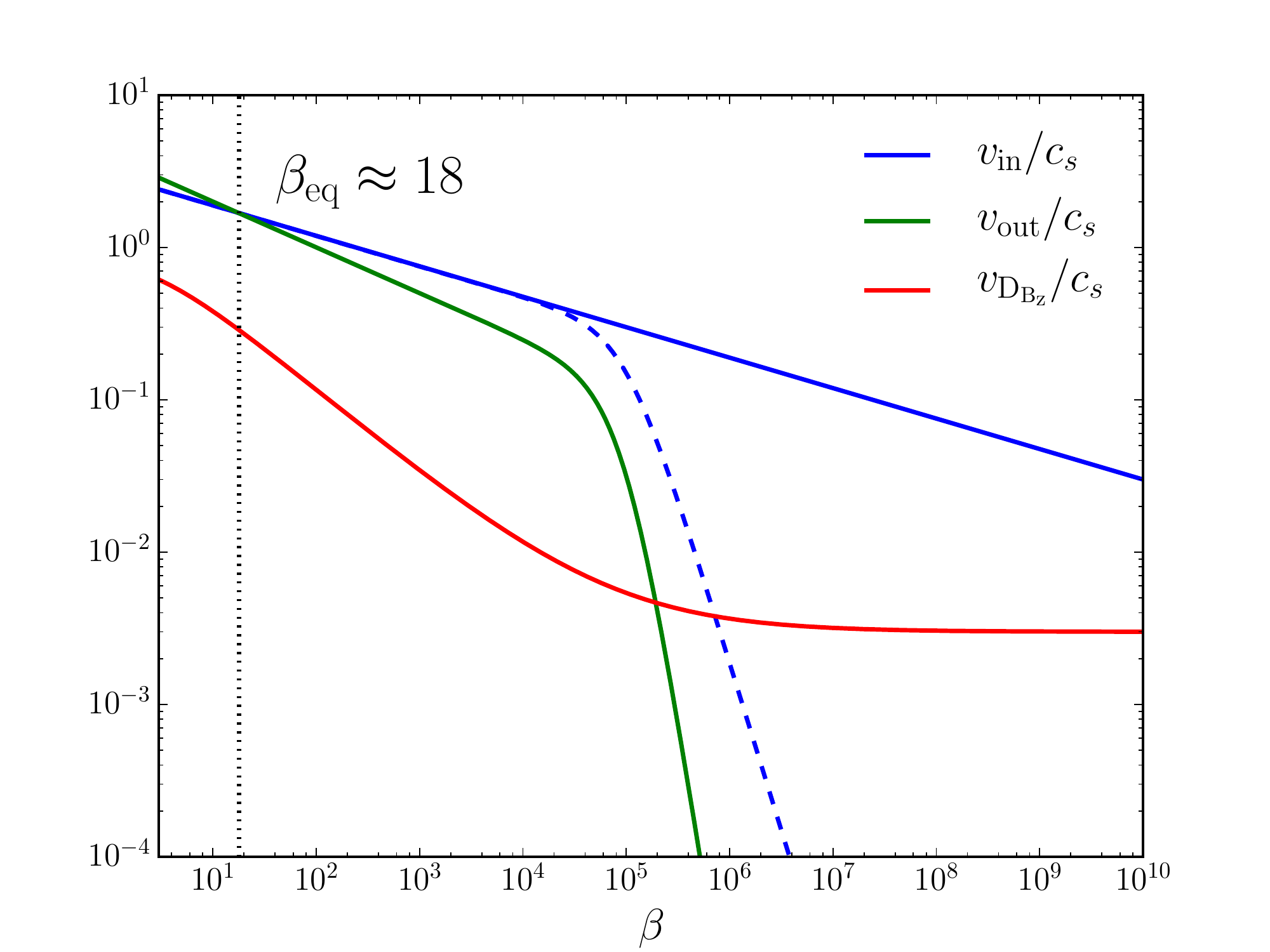}
\caption{Prescriptions for $v_\mathrm{in}$, $v_\mathrm{out}$ and $v_{D_{B_z}}$ normalized by the sound speed as a function of $\beta$. The solid blue line is supposed to represent a disk where the magnetic flux's advection by turbulent accretion is more efficient than diffusion at large $\beta$. The dashed blue line gives $v_\mathrm{in}(\beta)$ in a case where the magnetic flux's advection by turbulent accretion is less efficient than diffusion at large $\beta$ as studied in \S\ref{sec:noadvection}. The dotted black line gives the value of $\beta_\mathrm{eq}$.}
\label{fig:vpsi_pres}
\end{figure}

Figure \ref{fig:vpsi_pres} shows the prescriptions that we used. In this case $\beta_\mathrm{eq}\approx18$. Our results are not dependent on the exact value of $\beta_\mathrm{eq}$ as shown in Appendix \ref{sec:appendixB}.

\subsubsection{Transitory evolution}
\begin{figure*}[h!]
\begin{center}
\includegraphics[width=140mm,height=100mm]{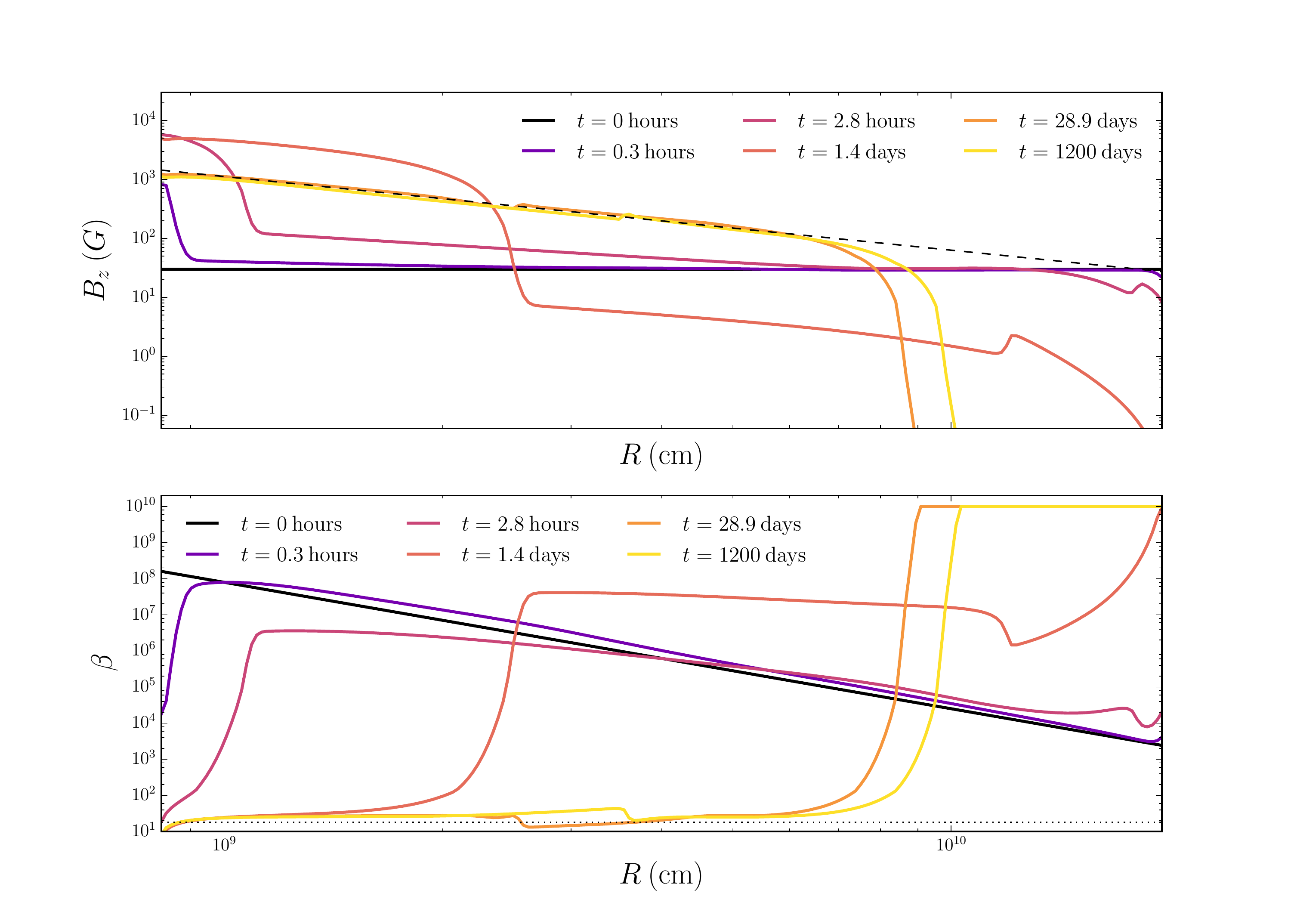}
\caption{Temporal evolution of the magnetic field configuration (upper panel) and plasma $\beta$ parameter (lower panel) up to the quasi-stationary final configuration. The initial magnetic field is uniform with a value of 30 G. The external mass transfer rate is $10^{16}$ $\mathrm{g\:cm^{-2}}$. The dashed black line in the upper panel gives the theoretical expected slope of the magnetic field, in $R^{-5/4}$ deduced from (\ref{eq:slope_Bz}). The dotted black line in the lower panel gives the value of $\beta_\mathrm{eq}$.}
\label{fig:Bz_evol}
\end{center}
\end{figure*}

In this section, we describe the early dynamics of the magnetic field when we evolve simultaneously the density, temperature and magnetic flux using the prescriptions from Figure \ref{fig:vpsi_pres}. We start from an initially uniform vertical large scale magnetic field of 30 G (top panel of Figure \ref{fig:Bz_evol}) and an external mass transfer rate of $10^{16}$ $\mathrm{g\:cm^{-2}}$. The lower panel of Figure \ref{fig:Bz_evol} shows the initial distribution of $\beta$. Initially $\beta$ decreases from $10^9$ at the inner radius to $10^4$ at the outer radius. Since $\beta>\beta_\mathrm{eq}$ the magnetic flux is advected inwards (see Figure \ref{fig:vpsi_pres}). 

The magnetic flux is advected inwards until the magnetic configuration reaches an equilibrium at $\beta=\beta_\mathrm{eq}$ where diffusion outwards ($v_\mathrm{out}$) compensates advection inwards ($v_\mathrm{in}$). We see from Figure \ref{fig:Bz_evol} that after $1.4$ days the inner regions have already reached the equilibrium magnetization while the outer disk still advects magnetic field inwards. As the magnetic flux is accumulated inwards the inner magnetically dominated region grows with time. At $t\approx30$ days the disk has almost reached a quasi-stationary state. Between, $t\approx30$ days and $t\approx1200$ days the disk is in a stationary state except for small fluctuations in all the quantities. The mass accretion rate is constant and the disk does not undergo thermal eruptions. This would correspond to the case of a nova-like. The inner region of the disk concentrates most of the magnetic flux and is stabilized at a $\beta$ slightly above $\beta_\mathrm{eq}$ (see lower panel of Figure \ref{fig:Bz_evol}). This small difference is due to the gradient of $B_z$ transporting magnetic flux outwards. The outer region of the disk is emptied of its magnetic flux and reaches the floor in $\beta$. We prescribed a ceiling at $\beta=10^{10}$ for numerical convenience but this does not affect our results.\\

\subsubsection{Time scales}\label{sec:time_scale}
We see from Figure \ref{fig:Bz_evol} that advection of the magnetic flux inwards is fast. The time scale on which the magnetic field evolves, $t_\mathrm{mag}$, is given by equation (\ref{eq:psi_adv_diff}) and is 
\begin{equation*}
t_\mathrm{mag}=\frac{R}{v_\psi}
\end{equation*}
In a highly magnetized disk where $\beta\approx1$ and the accretion is due to a surface torque from a magnetized outflow, $v_\mathrm{in}$ and $v_\mathrm{out}$ are expected to be of the order of the sound speed $c_s$ (see \S\ref{sec:est_vin} and \S\ref{sec:est_vout}) so that 
\begin{equation*}
t_\mathrm{mag}\approx\left(\frac{R}{H}\right)\frac{1}{\Omega}
\end{equation*}
when $\beta\approx1$. For example, at $R=10^{9}\:\mathrm{cm}$, $t_\mathrm{mag}\approx1\:\mathrm{h}$ at the beginning of our simulation, which is consistent with the time scales on Figure \ref{fig:Bz_evol}.

Moreover, the accretion velocity in this case is also $\approx c_s$, so that the accretion time scale, $t_\mathrm{acc}$ is also $\left(\frac{R}{H}\right)\frac{1}{\Omega}$. This is much smaller than the typical viscous time scale of an $\alpha$-disk, $t_\mathrm{vis}=\frac{1}{\alpha\Omega}(\frac{R}{H})^2$, which is of $\approx10$ years at the outer radius at the beginning of our simulation.\\

When $\beta\approx1$, we have $\alpha\approx 1$ and so the magnetic time scale and the accretion time scale are only an order $R/H$ larger than the thermal time scale, $t_\mathrm{th}\equiv1/(\alpha\Omega)$. In these conditions the magnetic configuration and the density can adapt very rapidly to the thermal structure. The classical thermal-viscous instability may exhibit a different behavior in these conditions.

\subsubsection{Magnetic configuration}\label{sec:mag_config}

The magnetic field, after a transition regime, maintains the same configuration from $t\approx 1$ day up until $t=1200$ days. The inner magnetic field follows a power law with an index of $R^{-5/4}$, and the outer magnetic field is almost null. We define $R_\mathrm{tr}$ as the transition radius between the inner magnetized region and the outer non-magnetized region; it also correspond to the radius where the magnetic field deviates from the power law in $-5/4$. When $R_\mathrm{tr}$ increases, we see that the value of the magnetic field at the inner boundary, $B_{z_\mathrm{in}}$, decreases.  

We can explain those results with a very simple model. The inner outflow-driven accretion zone has $\beta\approx\beta_\mathrm{eq}$, which is constant. In the inner zone, the accretion is driven by the magnetized outflow's torque and the mass accretion rate 
\begin{equation}\label{eq:mdotwind}
\dot{M}=qB_z^2\frac{R}{\Omega}
\end{equation}
is constant since the disk is steady. Since, $\beta$ in the inner zone remains fixed around $\beta_\mathrm{eq}$, $q(\beta)$ remains at $q(\beta_\mathrm{eq})$. Isolating $B_z$ in (\ref{eq:mdotwind}), we find
\begin{equation}\label{eq:slope_Bz}
B_z\propto R^{-5/4}
\end{equation}
in the inner zone as in \cite{blandford1982}. This simple analytical estimate is perfectly reproduced by our model as we can see from the black dashed line on the top panel of Figure \ref{fig:Bz_evol}.

\begin{figure*}[h!!]
\centering
\includegraphics[width=180mm,height=100mm]{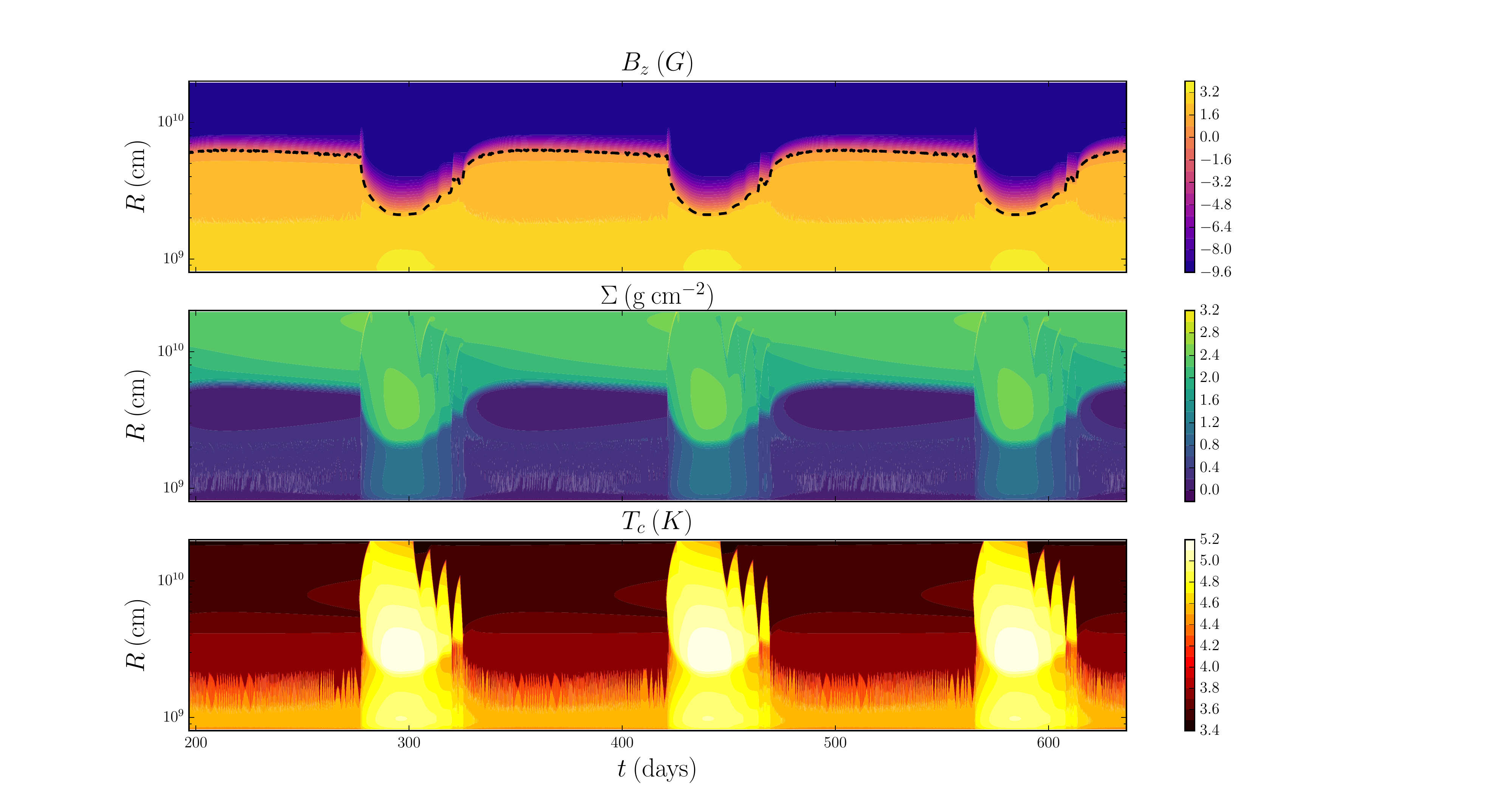}
\caption{Magnetic field (top panel), surface density (mid panel) and temperature (bottom panel) as a function of time and radius. The black dashed line in the top panel gives the analytical estimate of the transition radius using (\ref{eq:Rtrans}) and (\ref{eq:bz_in}). The initial magnetic field is uniform with $B_z=10$ G. The external mass transfer rate is $10^{16}$ $\mathrm{g\:cm^{-2}}$.}
\label{fig:bz_sigma_tc}
\end{figure*}

In an isolated disk ($v_\psi(R_\mathrm{out})=v_\psi(R_\mathrm{in})=0$), the magnetic flux is conserved. If we assume that all the magnetic flux is concentrated in the inner part, with $B_z\propto R^{-5/4}$ then the conservation of the magnetic flux
\begin{equation*}
\Psi=\int_{R_\mathrm{in}}^{R_\mathrm{tr}}B_z RdR,
\end{equation*}
where $\Psi$ is the total magnetic flux in the disk, can be rewritten as
\begin{equation}\label{eq:Rtrans}
R_\mathrm{tr}=R_\mathrm{in}\times\left(\frac{3}{4}\frac{\Psi}{B_{z_\mathrm{in}}R_\mathrm{in}^2}+1\right)^{4/3}.
\end{equation}
We can easily constrain the magnetic field at the inner boundary with (\ref{eq:mdotwind}) and we find
\begin{equation}\label{eq:bz_in}
B_{z_\mathrm{in}}=\left(\frac{\dot{M}_\mathrm{in}\Omega_\mathrm{in}}{qR_\mathrm{in}}\right)^{1/2}
\end{equation}
The values of $R_\mathrm{tr}$ and $B_{z_\mathrm{in}}$ are in perfect agreement with our simulations as we can see from the upper panel of Figure \ref{fig:bz_sigma_tc} where we report the temporal and radial evolution of the magnetic field with a dashed black line representing the analytical estimation (\ref{eq:Rtrans}) where the value of $B_{z_\mathrm{in}}$ was set by (\ref{eq:bz_in}). 

Let us finally point out that, in a disk where the magnetic flux is conserved and in the approximation that $R_\mathrm{tr}\gg R_\mathrm{in}$, the equations above imply that the transition radius evolves as
\begin{equation}\label{eq:rtr}
R_\mathrm{tr}\propto\dot{M}_\mathrm{in}^{-2/3}.
\end{equation}
This behavior is somewhat different from that of the magnetospheric radius $R_\mathrm{mag}$, set by equating the ram pressure of the accretion disk with the magnetic pressure from a possible dipolar magnetic field anchored in the white dwarf. Accretion then proceeds along magnetic field lines, truncating the disk within $R_\mathrm{mag}$. The magnetospheric radius behaves as $R_\mathrm{mag}\propto \dot{M}_\mathrm{in}^{-2/7}$ \citep{Frank}. We stress that in our case the disk is not truncated below $R_\mathrm{tr}$ but its properties are changed as it becomes highly magnetized. 

\subsubsection{Temporal evolution}\label{sec:temporal}

Figure \ref{fig:bz_sigma_tc} shows the evolution as a function of time and radius of the magnetic field, the surface density and the temperature. These results were found using an initial constant magnetic field of 3 G and a mass transfer rate of $10^{16}\:\mathrm{g\:cm^{-1}}$.

Figure \ref{fig:bz_sigma_tc} shows that the disk is unstable to the thermal-viscous instability. The only difference with the stable disk shown in Figure \ref{fig:Bz_evol} is a lower magnetic field ($B_z=10$ G) than in the stable case (30 G). In the high magnetic flux case of Figure \ref{fig:Bz_evol} the transition radius $R_\mathrm{tr}$ is larger than the radius where the thermal-viscous instability is triggered. This stabilizes the disk and allows it to find a stationary state with a constant mass accretion rate. We recover the instability by decreasing the magnetic flux in the disk. 

We see from Figure \ref{fig:bz_sigma_tc} that the eruptions are triggered at the transition between the inner magnetized and the outer non magnetized disk. Starting from the quiescent state, the disk accumulates mass until $\Sigma$ reaches a critical value where the disk becomes thermally unstable and transits to a hotter state, corresponding to the eruptive state. We see two fronts that propagates through the disk: one inwards and the other outwards. The disk stays in eruption for approximately 40 days. Then, a cooling front coming from the outer boundary of the disk signals the end of the outburst as it propagates inwards. We note that the disk has reflares \citep{menou1999} when going back to quiescence. These show up as oscillations in the size of the hot zone as the disk returns to quiescence (bottom panel of Figure \ref{fig:bz_sigma_tc}). This means that when the cooling front propagates the density reaches the critical density again and a new heating front appears. Then, another cooling front develops at the outer edge and, if the density in the disk is low enough, the cooling front can propagate all the way down to the transition radius ending the eruption. The disk then returns to quiescence and stays in quiescence for approximately 80 days. The recurrence of these eruptions depends on the initial magnetic field and the parameters of the disk as we will see in \S\ref{sec:LC}.\\

\begin{figure}[h!]
\begin{center}
\includegraphics[width=100mm,height=110mm]{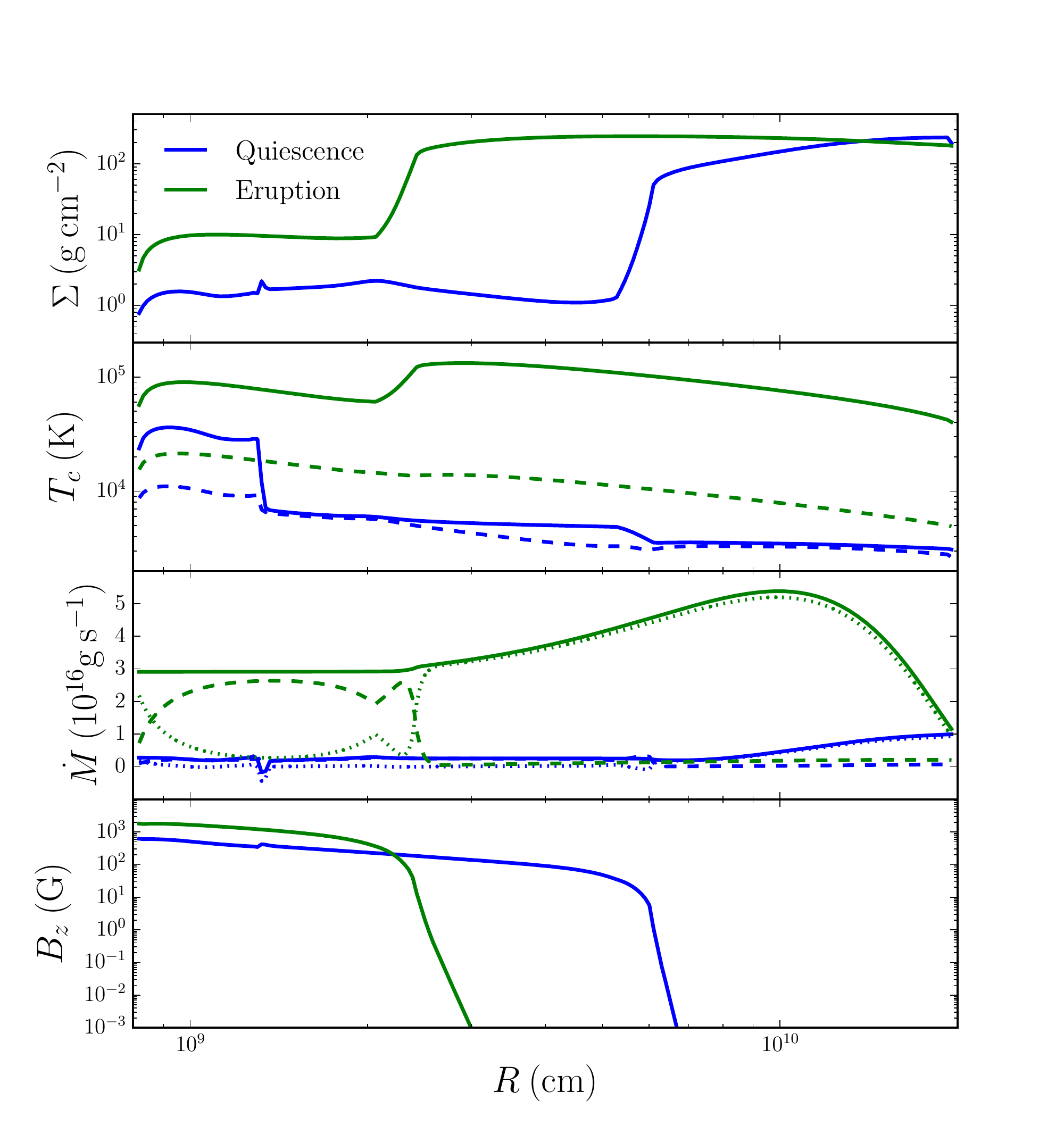}
\caption{Surface density (first panel), temperature (second panel), total mass accretion rate (third panel) and vertical magnetic field (fourth panel) as a function of radius in eruption (green curves) and quiescence (blue curves). The dashed curves in the second panel represent the effective temperature of the disk. The dashed curves in the third panel represent the outflow-driven mass accretion rate and the dotted curves represent the turbulent-driven mass accretion rate. The initial magnetic field is uniform with $B_z=10$ G.}
\label{fig:disk_config}
\end{center}
\end{figure}

From Figure \ref{fig:disk_config} we also see that the magnetic configuration changes between the eruptive state and the quiescent state. In fact, the evolution of the transition radius is driven by the thermal-viscous instability operating in the outer unmagnetized disk. We see that $R_\mathrm{tr}$ increases when going from eruption to quiescence in response to variations in $\dot{M}_{\rm in}$ (\ref{eq:rtr}). Going back to quiescence, $\beta$ at the transition decreases since the disk temperature drops. When $\beta$ decreases below $\beta_\mathrm{eq}$ the disk enters a zone where the diffusion is dominant, and the inner magnetic flux diffuses outwards until a $\beta=\beta_\mathrm{eq}$ is reached again. On the contrary, when going from quiescence to eruption, the value of $\beta$ increases. At the transition, the magnetic field is then advected inwards since $\beta$ becomes greater than $\beta_\mathrm{eq}$. The transition radius moves inwards as the heating front propagates. The evolution of the inner disk is entirely dictated by the eruptions in the outer disk.

\subsubsection{Radial structure}
Figure \ref{fig:disk_config} shows the disk's radial structure in quiescence and in eruption. The disk is radically different in the outer and inner parts during both eruption and quiescence. 

The outer disk is emptied of magnetic flux. The value of $\beta$ in this zone is set by the floor of $\beta=10^{10}$ that we impose artificially. Thus, the outer parts behave as a standard $\alpha$-disk with a constant $\alpha\approx0.03$. The accretion is due to turbulence and the accretion energy is deposited locally in the disk. The accretion driven by the torque of the magnetized outflow is very weak in this zone. It is in the outer disk that the thermal-viscous instability takes place. As we see from Figure \ref{fig:disk_config}, the outer disk in outburst is hot, dense, and optically thick. In quiescence, it is cold and has a lower density. 

The inner disk, in which most of the magnetic flux has accumulated, behaves very differently from an $\alpha$-disk. It reaches the value $\beta_\mathrm{eq}\approx18$ where accretion is driven by the torque from the magnetized outflow. Because this magnetized accretion is so efficient, i.e. $v_R\approx q/\beta c_s\geq c_s$, the density in the inner disk is low. A significant fraction of the accretion energy is not deposited locally but is lost in the outflow. Thus, the inner disk also has a lower temperature than a viscously-driven disk for the same $\dot{M}$ because of the lower heating rate. However, we find that the densities and temperatures in the inner disk are not as low as the values reached in \citet{scepi2019}, where the fixed magnetic field configuration led to very low values of $\beta$ in the inner disk (Figures 2 and 5 in \citealt{scepi2019}). Surprisingly, both $\Sigma$ and $T_{\rm c}$ remain high enough that the thermal-viscous cycle continues to operate in the inner disk.  Figure \ref{fig:disk_config} also show the presence of a small hot zone in $T_{\rm c}$ in the magnetized inner disk whereas the viscously-driven outer disk is in quiescence. This produces small eruptions on a very short timescale, because of the fast accretion timescale in the magnetized inner disk. These eruptions shows up as wiggles during quiescence in panels (c) and (d) of Figure \ref{fig:mag}.

\subsubsection{Light curves}\label{sec:LC}
\begin{figure}[h!]
\includegraphics[width=100mm,height=100mm]{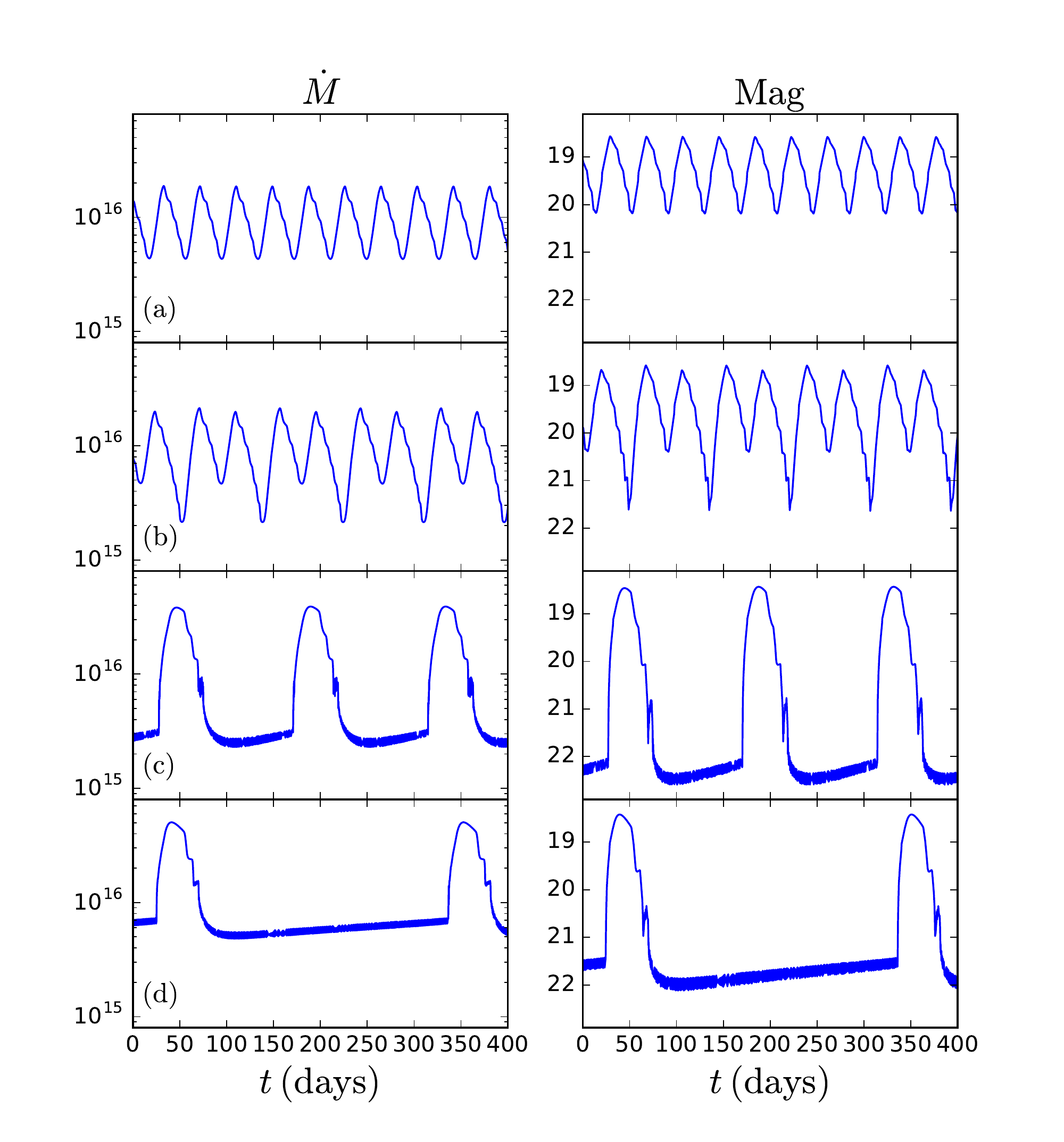}
\caption{Mass accretion rate at the inner boundary (left panels) and V magnitude (right panels) as a function of days for different magnetic flux. From top to bottom the initial constant magnetic field $B_{z_0}$ is equal to 0.1, 1, 10 and 21 G.}
\label{fig:mag}
\end{figure}

We showed in \S\ref{sec:temporal} that the magnetic field configuration is self-consistently changing from the eruptive state to the quiescent state. As we can see from Figure \ref{fig:mag}, by changing the value of the initial magnetic flux we can change the shape of the eruptions. 

For a weak initial magnetic field ($B_{z_0}=0.1\:$G) accretion driven by the magnetized outflow is negligible as there is not enough magnetic flux in the disk. The disk is mainly turbulent with $\alpha\approx0.03$ and the eruptions are of small amplitude with short recurrent time scale (panel (a) of Figure \ref{fig:mag}). We attribute the differences in amplitude and recurrence timescale with the low magnetic field case shown in the top panel of Figures 3-4 in \citet{scepi2019}, where the same disk is modeled, to our simplifying assumptions (\S\ref{sec:DIM}). As we increase the magnetic flux in the disk, the accretion driven by the magnetized outflow starts to be important in the inner disk. The transition radius   acts as an effective truncation since the accretion timescale becomes very short below $R_\mathrm{tr}$. Indeed, Figure \ref{fig:mag} shows the lightcurve behaves as expected from a disk with a progressively higher truncation radius as the magnetic flux is increased. The heating front that triggers the outburst starts close to $R_\mathrm{tr}$ for these parameters. Since the critical surface density for the ignition of the viscous-thermal instability is proportional to $R^{1.1}$  \citep{hameury1998}, a truncated disk needs to build more mass between outbursts, increasing the recurrence time scale from panels (a) to (c). The mass accretion rate in quiescence also increases as $R_\mathrm{tr}$ increases (left panels of Figure \ref{fig:mag}), an effect that has long been identified as a possible solution to the high X-ray flux measured from DNe in quiescence (\citealt{mukai2017} and references therein). The disk becomes `leaky' when its $\dot{M}_{\rm in}$ in quiescence becomes close to the mass transfer rate \citep{lasota2001}, preventing mass from building up in the disk and further increasing the recurrence time scale as illustrated when going from panels (c) to (d).
These results do not depend strongly on the exact value of $\beta_\mathrm{eq}$. We retrieved similar light curves with different $\beta_\mathrm{eq}$ by changing only the amount of magnetic flux in the disk (see Appendix \ref{sec:appendixB}).

\subsection{Inefficient advection at large $\beta$}\label{sec:noadvection}
We now briefly explore the second case where the advection of magnetic flux at large $\beta$ is inefficient. This case aims at mimicking the results of \cite{lubow1994}. We use the following expressions for $v_\mathrm{in}$ :
\begin{equation}
v_\mathrm{in}=3f\times\beta^{-0.2}
\end{equation}
where $f=1/(1+10^{-10}\beta)$. $v_\mathrm{out}$ is still defined according to (\ref{eq:vout}). The only difference with our previous prescription is that $v_\mathrm{in}$ is also quenched at large $\beta$ as shown by the dashed blue line in Figure \ref{fig:vpsi_pres}. Hence there exists a magnetization above which the magnetic flux cannot be advected anymore because $v_\mathrm{in}<v_{D_{B_z}}$ (see Figure \ref{fig:vpsi_pres}). In \cite{lubow1994} the dominant term in the diffusion of the magnetic flux is due to the vertical diffusion of the radial magnetic field, where the radial magnetic field is computed from the global magnetic structure. Our model is local and is unable to capture such an effect. We use a case where $v_\mathrm{in}<v_{D_{B_z}}$ as a proxy for the global method of \cite{lubow1994}.\\

\begin{figure}[h]
\includegraphics[width=90mm,height=90mm]{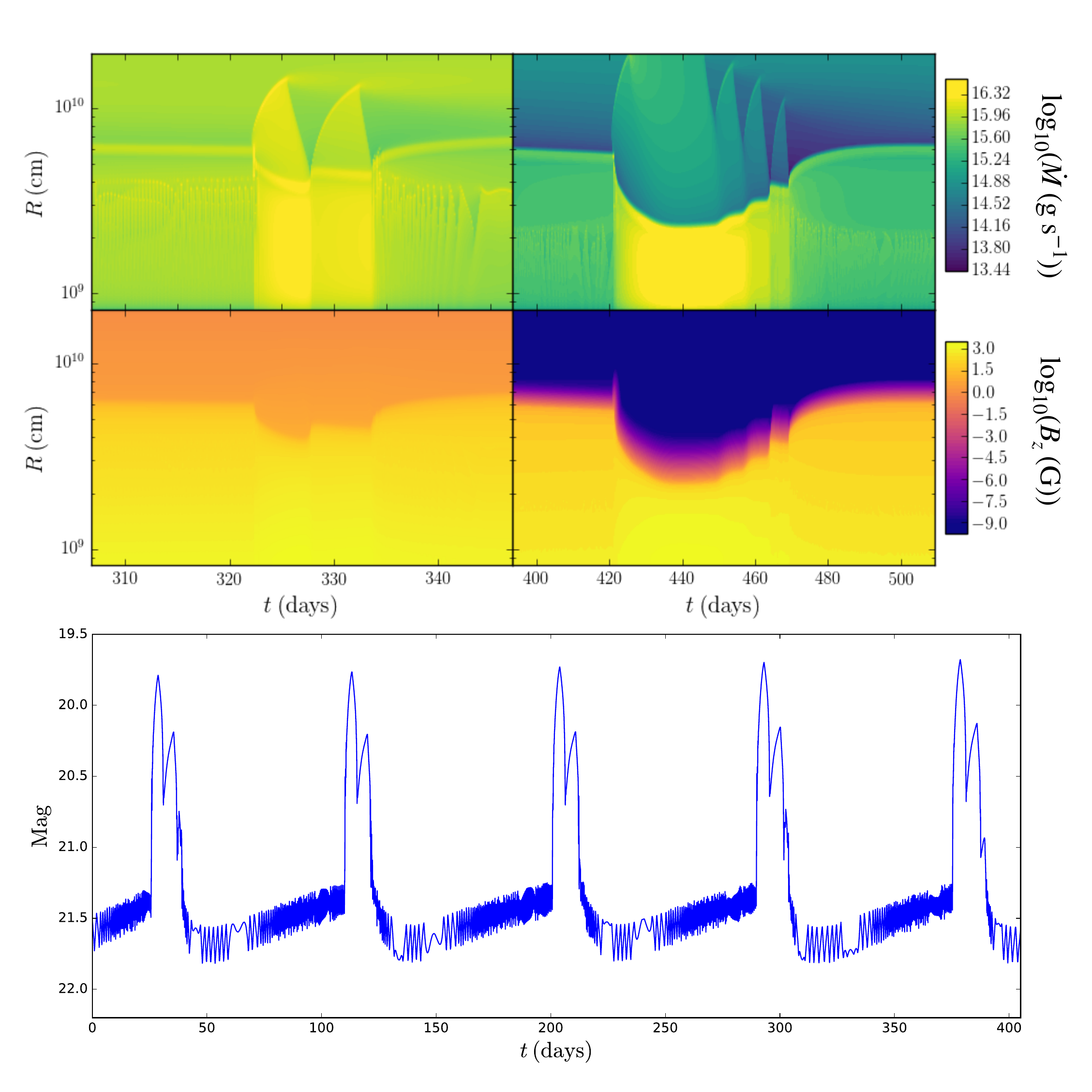}
\caption{The upper figure gives the temporal and radial evolution of the magnetic field (upper panels) and  of the outflow-driven mass accretion rate (lower panels) for the case of inefficient advection of the magnetic flux at high $\beta$ (left hand side panels) and for the case of efficient advection of the magnetic flux at high $\beta$ (right hand side panels). The lower figure shows the light curves for inefficient advection at large $\beta$. The initial constant magnetic field in this case is $B_{z_0}=20$ G.}
\label{fig:LC_inefficient}
\end{figure}

We do not discuss this case in as much detail as the one of efficient advection at large $\beta$. Indeed, the results are very similar except that the light curves obtained with this prescriptions show eruptions of smaller amplitude ($\approx$ two magnitudes) as can be seen in Figure \ref{fig:LC_inefficient}. This is due to the fact that the magnetic field at the inner boundary cannot change as much between the eruptive state and the quiescent state as in the case of efficient advection (see a comparison between both cases on Figure \ref{fig:LC_inefficient}). In the case of inefficient advection, the disk can only advect inwards (or diffuse outwards) the magnetic flux from the parts of the disk where $\beta\lesssim10^5$. This leaves the outer parts of the disk out of reach and limits the range over which the magnetic field at the inner boundary can vary. Ultimately, this limits the range over which the transition radius can change (see Figure \ref{fig:LC_inefficient}) and produces smaller eruptions than in \S\ref{sec:LC}.

\section{Discussion}\label{sec:discussion}

\subsection{Limitations of our model}\label{sec:limitations}
Our model assumes that the vertical velocity and the radial magnetic field are small enough so that we can neglect the term $v_zB_R$ compared to $v_RB_z$ in the induction equation. However, the velocity $v_z$ can be important, especially in the inner magnetized region where a wind or a jet is formed. In the case where $\beta_\mathrm{eq}$ is $1-100$, $v_z$ becomes $\gtrsim 2$ times the local orbital speed \citep{jacquemin2019} and so $v_Z B_R/v_R B_Z$ cannot be neglected anymore since $B_R/B_Z\gtrsim1$ and $v_R\lesssim c_s$. This could induce another source of magnetic flux transport.

Moreover, we assume that there exists an outflow transporting angular momentum but we do not treat the transport of mass and energy in the outflow. This means that the disk may be even emptier and cooler than it is in the magnetized outflow-dominated inner parts on Figure \ref{fig:bz_sigma_tc}.\\ 

Also, our approach is different from the one of \cite{lubow1994}, \cite{guilet2014} or \cite{li2019} in that their model is global whereas our approach is purely local. We assume that the behavior of the disk only depends on the local magnetization and not on the full global structure. We note that the prescriptions from \cite{jacquemin2019} are self-similar and so, by nature, take into account the global structure of the disk. However, due to the self-similar approximation the solutions from \cite{jacquemin2019} have a uniform magnetization (not dissimilar to global simulations from \citealt{zhu2018} and \citealt{mishra2020}) and are not consistent with a realistic disk where the magnetization depends on the radius and is not self-similar. This is especially true close to the fronts.

We would like to emphasize that the global approach followed by \cite{lubow1994}, \cite{guilet2014} and \cite{li2019} relies on the approximation that the magnetic field is potential. When an outflow is present this is clearly not true because the outflow matter's inertia exerts a force on the magnetic field configuration. Our local approximation may be more suited than the potential approximation when an outflow is present though it still needs a proper support from global simulations. Nevertheless, the diffusion terms in the outer disk, which has a high $\beta$ and no outflow, may be underestimated as we do not take into account the global structure.\\

At high magnetizations ($\beta\lesssim10^2$) other instabilities that we ignored here such as the magnetic Rayleigh-Taylor instability might play a role in giving the final magnetic configuration as seen in simulations of \cite{mckinney2012}. Taking into account these high magnetization instabilities would require a detailed study providing a simple prescription of the mass, angular momentum (see \citealt{marshall2018}), energy and magnetic field transport that we can use in our model. However, it is not clear that a simple axisymetric model is be able to reproduce the long term behavior of high magnetizations simulations, which are reported to be highly non-axisymetric with spiral patterns possibly related to the properties of transport (\citealt{mckinney2012},  \citealt{mishra2020}).

\cite{beckwith2009} also expressed concerns regarding the use of a mean field approach to treat the turbulence. In their 3D GRMHD global simulations, turbulent motions reconnect field lines in the surface accretion layer forming closed field loops in the disk and allowing coronal advection of open magnetic field lines on the horizon of the central black hole. A detailed study of whether this effect can be captured or not using a mean field approach remains to be done.\\

Finally, we restricted our model to a case where the magnetic flux is conserved. It is not clear whether such a case is relevant for compact binaries where the disk is fed with matter and possibly magnetic flux at the outer radii. We tried to address this problem by feeding continuously the disk with net vertical magnetic field. However, in the absence of any mechanism to extract magnetic flux from the disk, the magnetic flux keeps accumulating until we stop the simulation. To prevent a dramatic accumulation of magnetic flux, we also tried to feed the disk with a vertical field of switching polarity. In the case of efficient advection at large $\beta$, we mostly find modulations of the light curves with an amplitude and a frequency corresponding to those of the injected magnetic flux. It seems unlikely that the eruptions of DNe should actually be driven by the varying incoming magnetic field and so we did not pursue in this direction. Internal local generation of net vertical magnetic field by the MRI dynamo might also play a role in the long term evolution of DNe (see \citealt{begelman2014} for low mass X-ray binaries). However, exploring such effects is clearly out of the scope of this paper.\\

\subsection{Comparison with \cite{scepi2019}}
One major result of this paper is that the inner disk, where accretion is driven by the torque due to the magnetized outflow, acts exactly as a truncation of the inner disk. The inner disk is passive and responds to the changes in the outer disk; it does not drive the dynamics of the eruptions. Our hybrid disk can then simply be modeled as a truncated $\alpha$-disk with a varying inner radius, $R_\mathrm{in}$ that follows the relations (\ref{eq:Rtrans}) and (\ref{eq:bz_in}). To check this assertion, we performed a simulation of such a truncated $\alpha$-disk and found the same behavior than reported in Figure \ref{fig:bz_sigma_tc}.

This is different from the results of \cite{scepi2019}, where the inner zone acted only partly as a truncation of the disk.  Because the magnetic field configuration had a fixed dipolar configuration, the transition in outflow-driven mass accretion rate was more shallow than in the present work. Hence, the magnetized outflow's torque also enhanced the accretion in the outer disk (especially during outburst) and did not act as a simple truncation. Indeed, we had checked in \cite{scepi2019} that a simple truncation of the outer $\alpha$-disk led to weak outburst with short recurrence time scales. Here, the magnetic flux is much more concentrated to the disk center and the transition between magnetized and unmagnetized regions is sharper.

Another difference is that the relation between the transition radius, $R_\mathrm{tr}$, and the mass accretion rate reported here is different from the one of \cite{scepi2019}. In the case of a fixed dipolar magnetic field, $R_\mathrm{tr}$ evolves as $\dot{M}^{-2/7}$ (like the magnetospheric radius) while here it evolves as $\dot{M}^{-2/3}$ (\ref{eq:rtr}). Since a disk can be stabilized if $R_\mathrm{tr}>R_\mathrm{inst}$, where $R_\mathrm{inst}$ is the radius at which the thermal instability would be triggered, this modifies the dependency with orbital period of the stability criterion shown on the Figure 6 of \cite{scepi2019}, for a given amount of magnetic flux.
 
\subsection{Comparison with global simulations}

Because of the difference of time scale, it is difficult to directly compare our work with 3D global (GR-)MHD simulations. We are interested here in the secular evolution, i.e. over several viscous time scale while state-of-the-art 3D global simulations could typically only reach hundreds of orbital time scales at the outer radius of DNe. Still, \cite{liska2018b} performed a long duration GRMHD 3D global simulation, and found that at late times the accretion disk is divided into an inner magnetized, low density region and an outer region sharing the properties of an $\alpha-$disk. The authors suggested that a large scale magnetic torque might be responsible for the formation of this inner region. 

More generally GRMHD simulations studying magnetic flux transport report that the magnetic flux is advected very efficiently inwards (\citealt{beckwith2009}, \citealt{mckinney2012}). These simulations typically start with a torus and let the accretion disk form. However, the most recent MHD simulations, which initially start with an accretion disk with a constant $\beta$, tend to show that the magnetic flux is neither advecting inwards or outwards (\citealt{zhu2018}, \citealt{mishra2020}), in tension with the GRMHD results. This result, surprisingly, seems to hold whatever the value of $\beta$ between $10^2$ and $10^4$. This conflict between GRMHD and MHD simulations clearly deserves further attention from the community.

\subsection{Comparison with observations}
Letting the magnetic field distribution evolve under the assumption of efficient advection concentrates the flux in an inner magnetized disk. The outer disk is purely viscous and behaves dynamically as if it were truncated at the transition radius between the two zones. This has several  observational consequences.

First, this provides an alternative mechanism through which disk truncation can be achieved. Disk truncation has been invoked to explain the delayed rise to outburst of the UV flux, compared to the optical flux, and the relatively high accretion rate onto the white dwarf inferred from the X-ray observations in quiescence, compared to the predictions from the DIM for a disk extending down to the surface of the white dwarf (see \citealt{lasota2001,mukai2017} and references therein). Various mechanisms have been proposed: truncation by the white dwarf magnetic field as in intermediate polars \citep{1992MNRAS.259P..23L}, evaporation of the inner disk by a `coronal siphon flow' \citep{1994A&A...288..175M}, white dwarf irradiation keeping the inner disk hot \citep{1997MNRAS.288L..16K}. We propose that it is the magnetic field carried by the disk itself that causes an apparent truncation of the disk. Such a mechanism would apply equally well to low-mass X-ray binaries as we discuss further below.

Second, the truncation of the inner disk by the outflow-driven accretion will shape the light curve. The eruptions have a longer recurrence time scale and a larger amplitude than in the case of a purely turbulent disk with a constant $\alpha$  (Figure \ref{fig:mag}). \citet{coleman2016} argued that truncation of the inner disk was essential to match outburst models, using the actual values of $\alpha$ measured in simulations of MRI turbulent transport, to observations. They proposed truncation by the white dwarf dipolar field. Our model could produce such a truncation without requiring every DNe to have strong magnetic fields on their white dwarf. We also predict a different evolution of the truncation radius than for the magnetospheric radius. 

Third, the magnetic flux in the disk will be one of the two main parameters, with the external mass transfer rate, shaping the lightcurve. Observationally, there exists a large scatter in the recurrence time scale of DNe \citep{cannizzo1988} that is usually explained by external mass transfer rate variations at a given orbital period. In our model, we could naturally account for the dispersion in recurrence time scales by varying the initial magnetic flux in the disk. We defer to future work a more detailed comparison of lightcurves to observations using (\ref{eq:rtr}) to evolve the truncation radius, together with a more realistic treatment of the radiative transfer of the thermodynamics of the disk \citep{hameury1998}.

Fourth, a high enough magnetic flux will also affect stable systems. It can stabilize a disk that would be unstable to the classical thermal-viscous instability by truncating the inner hot region, leaving only a stable outer cold region, much like posited for magnetospheric truncation (\citealt{lasota1995}, \citealt{menou1999b}). This would produce a population of cold, stable systems. Hot, stable systems (nov\ae\ likes) can also be impacted.
Actually, in nov\ae-like variables, there exists a discrepancy between the observed spectrum and the spectrum expected from a classical $\alpha$-disk extending to the inner radius. On the one hand, the emission coming from the inner radii of the disk seems to be missing (see \citealt{nixon2019} for recent work and review on the subject). On the other hand, dwarf nov\ae\ in eruption, whose observational properties are expected to be close to those of nov\ae\ likes, are well fitted by an $\alpha$-disk extending to the inner radius \citep{hamilton2007}. The discrepancy could simply arise because the disk, for the same mass accretion rate, can be more or less truncated depending on the available magnetic flux.

Last, an observational signature of our model is the presence of magnetized outflows, jets or winds. Recent observations of radio emission in DNe and nov\ae-like have been attributed to magnetic jets (\citealt{kording2008}, \citealt{russell2016}) though more than one mechanism could produce such emission (\citealt{coppejans2015}, \citealt{coppejans2016}). If magnetic jets are indeed responsible for the radio emission, we could expect from our model a correlation between strong radio emitters and DNe with long-recurrence time scale. Also, since DNe with a large magnetic flux are stabilized we could also expect to find cold, stable systems with strong radio emission. Matching stable, point-like sources in radio surveys to late-type stars in optical surveys may be a way to identify those optically-faint systems.

\subsection{Application to LMXBs}\label{sec:LMXBs}
Low mass X-ray binaries are analogues of DNe where, instead of a white dwarf, the central object is a stellar mass black hole or a neutron star. The spectral and variability properties of LMXBs provide several lines of evidence for a truncated thin disk (see \citealt{2014ARA&A..52..529Y} for a review and \citealt{2020ApJ...896L..36Z} for more recent work). These objects have eruptions that can also be explained by the DIM  taking into account the X-ray irradiation of the outer region by the inner region and a truncated disk \citep{1996A&A...314..813L,dubus2001}. 

\cite{ferreira2006} suggested (and \citealt{marcel2018a,marcel2018b} later showed) that all the spectral states of LMXBs could be reproduced using an hybrid disk composed of : 1) an inner jet-emitting disk (JED), with $\beta\approx1$, where all the angular momentum and part of the energy are transported vertically by a magnetic jet ; 2) a standard $\alpha$ accretion disk (SAD). By varying only the inner mass accretion rate and the transition radius between the JED and the SAD, \cite{marcel2019} were able to reproduce the outburst of GX 339-4.

Our disk configuration is not very different from what has been assumed by \cite{ferreira2006} and \cite{marcel2019}. The main difference between the inner JED disk and our inner magnetized outflow-driven accretion dominated zone is the degree of magnetization. While the JED solutions require $\beta\approx1-10$ we have a $\beta$ that can range from $10$ to $10^3$ in our model without affecting the light curves (see Appendix \ref{sec:appendixB}). The SAD, outer disk of \cite{marcel2019} is our outer, non-magnetized, turbulent-driven accretion disk. A hybrid disk with a sharp transition between these two components being a natural outcome of our model, we propose that our model might provide a dynamical justification to the model of \cite{ferreira2006} and might be able to explain part of the spectral evolution of LMXBs. This will be the subject of future work.
 
\section{Conclusions}\label{sec:conclusion}
We studied the evolution of a non-steady, axisymmetric, DNe accretion disk undergoing eruptions due to the thermal-viscous instability. The novelty of this work is that we evolve the large scale poloidal magnetic configuration during the outburst cycle using a local toy model for magnetic flux transport. The disk is isolated in the sense that it cannot lose or gain magnetic flux. To evolve the density, temperature and magnetic field, we developed a version of the classical DIM \citep{hameury1998} that takes into account the removal of angular momentum by MRI turbulence and by magneto-centrifugal outflows as in \cite{scepi2019}.  We assumed that the magnetic flux evolution only depends on the local magnetization and used several different sets of prescriptions for magnetic flux transport to study their impact on the outburst light curves.

The only case where we find realistic light curves is when magnetic flux is advected inwards until it reaches an equilibrium magnetization $\beta_\mathrm{eq}$ for which inward magnetic field advection is balanced by outward magnetic field diffusion. This case naturally leads to a hybrid disk configuration, regardless of the exact value of $\beta_\mathrm{eq}$. The inner region of the disk is highly magnetized and has a uniform $\beta(R)=\beta_\mathrm{eq}$. As a consequence, the magnetic field is distributed as $R^{-5/4}$ in this region. The angular momentum and accretion energy are mostly transported vertically in the outflow. The outer region of the disk is emptied of magnetic field thus outflow-driven accretion is negligible there. The outer disk resembles a classical, viscously-driven, $\alpha$-disk.  

The transition between the inner outflow-dominated region and the outer turbulent region is very sharp and is of the order of the typical scale height of the disk. 
We find that our hybrid disk can actually be reduced to a truncated $\alpha$-disk with an inner boundary evolving as $\dot{M_\mathrm{in}}^{-2/3}$, which can be easily implemented in DIM codes. Thus, our model provides physical justification to previous models requiring, or observations suggesting, a truncation of the viscous disk in both DNe and LMXBs (see \citealt{hameury2019} and references therein). Here, the inner regions are replaced by a disk with a very fast accretion timescale driven by a strongly-magnetized outflow.

\begin{acknowledgement}
We thank Jean-Pierre Lasota, the referee, for a thorough and critical report that led to a better understanding and presentation of our results. NS would like to thank Jer\^{o}me Guilet for help at the beginning of this project, Gr\'{e}goire Marcel for many discussions and sharing unpublished results as well as Mitch Begelman and Jason Dexter for reading an early version of this work. NS acknowledges partial financial support from the p\^ole PAGE of the Universit\'e Grenoble Alpes and from a NASA Astrophysics Theory Program grant NNX16AI40G. Some of the computations presented in this paper were performed using the Froggy platform of the CIMENT infrastructure (https://ciment.ujf-grenoble.fr), which is supported by the Rh\^one-Alpes region (GRANT CPER07\_13 CIRA), the OSUG@2020 labex (reference ANR10 LABX56) and the Equip@Meso project (reference ANR-10-EQPX-29-01) of the programme Investissements d'Avenir supervised by the Agence Nationale pour la Recherche. GL acknowledges support from the European Research Council (ERC) under the European Union’s Horizon 2020 research and innovation programme (Grant agreement No. 815559 (MHDiscs)). GD acknowledges support from CNES.
\end{acknowledgement}

\bibliographystyle{aa}
\bibliography{biblio}

\begin{appendix}
\section{$v_\psi$ from \cite{guilet2012}}\label{sec:appendixA}
In a first approach, we tried to compute $v_\psi$ as a function of $\beta$ by obtaining the vertical structure of $v_R$ from the asymptotic expansion of \cite{guilet2012} instead of simple power laws. As mentioned in \cite{guilet2012}, there is a tight connection between solving the vertical structure of the disk in the thin disk asymptotic expansion and solving the eigenvalue problem for viscous, resistive MRI. Indeed, the vertical structure is a linear combination of MRI modes. The growth rates and the eigenvectors of the resistive MRI modes depends on the resistivity and viscosity and so ultimately on $\alpha$. \cite{guilet2012} used the value of $\alpha$ necessary to damp all MRI unstable modes to find $v_\psi$. However, this leads to unrealistic values, for example $\alpha\approx0.7$ for $\beta\approx10^5$. In our case, we use the value of $\alpha$ from 3D local MRI simulations with radiative transfer from \cite{scepi2018b}. Doing so, we do not ensure that all MRI modes are stable for the whole range of $\beta$ we considered. In fact, it is natural that there exists unstable MRI modes with our choice of $\alpha$ since the simulations we are based on are fully turbulent. When we use $\mathcal{P}=1$ we find that, near the values of $\beta$ where resistive MRI modes are damped by resistivity, our problem is ill-conditioned and we cannot find one unique solution. This is fundamentally due to the fact that it is physically difficult to find a stationary linear solution to a system that still has unstable modes.

We overcame that problem by enhancing the resistivity to damp all the unstable MRI modes. We found that for $\mathcal{P}\lesssim0.01$ $v_\psi$ is well behaved as a function of $\beta$. However, depending on the exact value of $\mathcal{P}$ we can be in different scenarii of Figure \ref{fig:vpsi_scenarii}. Then, given the uncertainty of our method, we switched to a more simplified method using simple power laws allowing us to probe a larger range of behavior.

\section{Light curve dependance on $\beta_\mathrm{eq}$}\label{sec:appendixB}
\begin{figure}[h!]
\includegraphics[width=90mm,height=100mm]{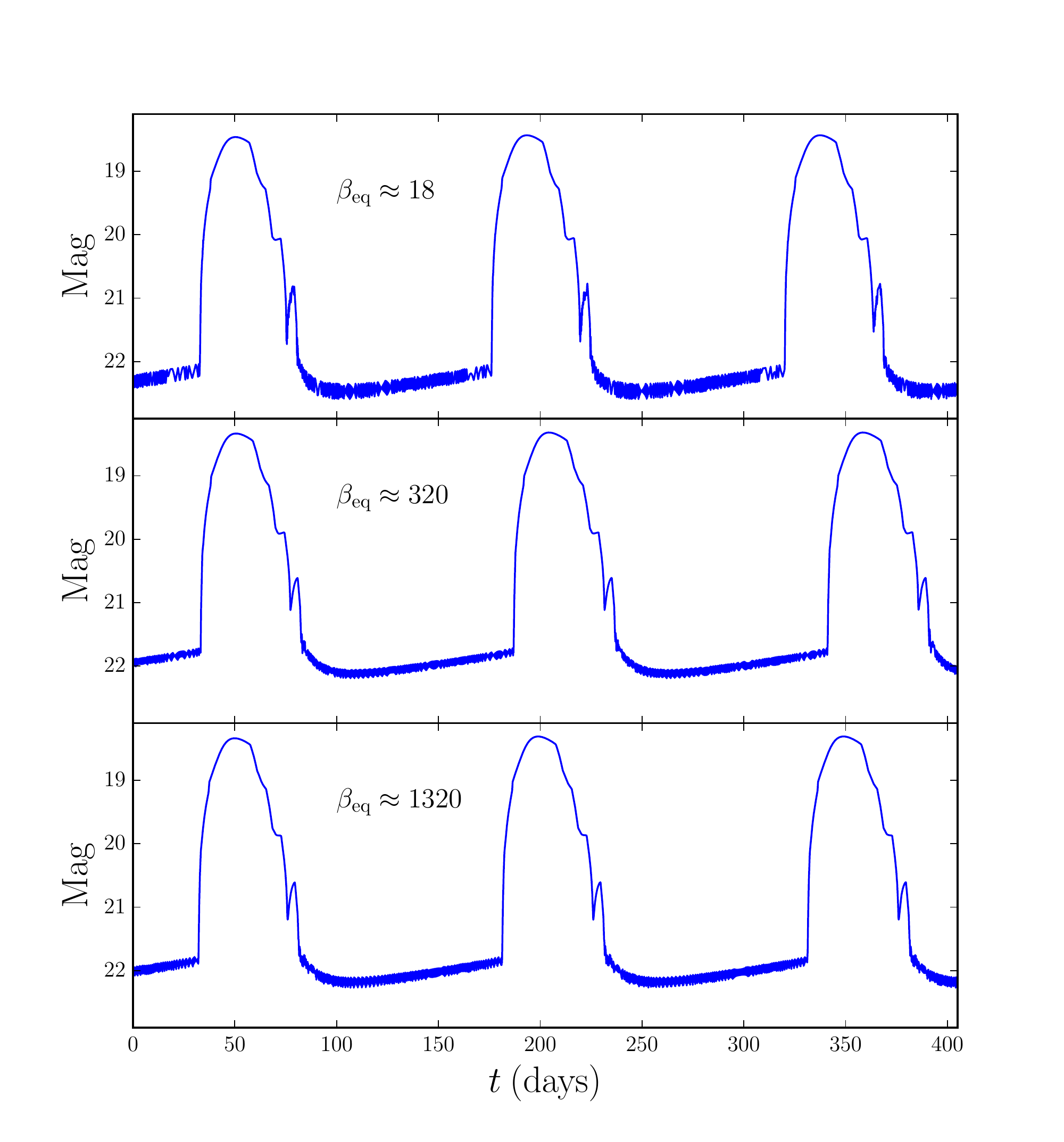}
\caption{Dwarf nov\ae\ light curves as a function of time for different values of $\beta_\mathrm{eq}$. We used $B_{z_0}$= 10, 4 and 2.5 G for the panels (a), (b) and (c) respectively.}
\label{fig:LC_prescriptions}
\end{figure}

The purpose of this paper is to show which parameters are important to have realistic dwarf nov\ae\ light curves. In this Appendix, we change the value of $\beta_\mathrm{eq}$ by slightly changing the slope of $v_\mathrm{out}$. We use three different values of $\beta_\mathrm{eq}$: 18, 320, 1320. We find that the exact choice of $\beta_\mathrm{eq}$ do not affect the shape of the light curves as can be seen from Figure \ref{fig:LC_prescriptions}. Indeed, by adjusting the initial magnetic flux we can find in each case light curves that look alike one another. Of course as $\beta_\mathrm{eq}$ becomes large the magnetized outflow's torque becomes less important changing the light curves. However, as long as $\beta_\mathrm{eq}$ is $\lesssim1000-10000$ the exact value is not crucial to the overall shape of the light curves.

\section{An outer magnetized disk}\label{sec:appendixC}
In the fourth case of Figure \ref{fig:vpsi_scenarii}, $v_\mathrm{out}>v_\mathrm{in}$ for $\beta>\beta_\mathrm{eq}$. Consequently, we could imagine the opposite magnetic field configuration compared to \S\ref{sec:efficient}: an outer magnetized disk and an inner weakly magnetized disk. As in \S\ref{sec:mag_config}, the outer magnetized disk would have a constant magnetization $\beta_\mathrm{eq}$ and thus a magnetic field distributed as $B_z\propto R^{-5/4}$. Again, similarly to \S\ref{sec:mag_config} we have the following constraint on the outer magnetic field:
\begin{equation}\label{eq:Bz_Rout}
B_{z_\mathrm{out}}=\left(\frac{\dot{M}_\mathrm{out}\Omega_\mathrm{out}}{qR_\mathrm{out}}\right)^{1/2}.
\end{equation}
If $B_{z_\mathrm{out}}$ is too large then $\dot{M}(R_\mathrm{out})$ becomes larger than the mass transfer rate from the companion, $\dot{M}_\mathrm{ext}$. In this case, the outer disk empties leading to a decrease in $\beta$. The outer disk then falls in the regime $\beta<\beta_\mathrm{eq}$ causing the magnetic flux to be advected inwards to return to the equilibrium magnetization $\beta_\mathrm{eq}$. However, the outer region acts as an enormous reservoir of magnetic flux. A small magnetic field advected inwards becomes a large magnetic field in the inner region. This also means that as one advects magnetic field inwards the outer magnetic field (and so the mass accretion rate) does not vary much. Thus, the disk keeps emptying and ultimately becomes strongly magnetized everywhere. In this case there is no eruptions.

To avoid this situation one needs $\dot{M}(R_\mathrm{out})$ to be of the order of $\dot{M}_\mathrm{ext}$. For $\dot{M}_\mathrm{ext}\approx10^{16}\:\mathrm{g\:s^{-1}}$ one can find using (\ref{eq:Bz_Rout}) that $B_{z_\mathrm{out}}\lesssim3$ G, where we have used $q\approx1$ to set an upper limit for the outer magnetic field. Setting $\Sigma\approx10^2\:\mathrm{g\:cm^{-1}}$, $H\approx10^8\:\mathrm{cm}$, one must have $\beta_\mathrm{eq}\gtrsim10^4$ to satisfy $\dot{M}(R_\mathrm{out})\approx\dot{M}_\mathrm{ext}$. We tried a simulation with $\beta_\mathrm{eq}=10^5$ and $\dot{M}_\mathrm{ext}=10^{16}\:\mathrm{g\:s^{-1}}$ and find small amplitude eruptions ($\approx2$ magnitudes) with very short recurrence time scales ($\approx1$ day).

One could also imagine to increase $\dot{M}_\mathrm{ext}$ to allow lower values of $\beta_\mathrm{eq}$. However, we are limited by the fact that the disk needs to be thermally unstable. If $\dot{M}_\mathrm{ext}$ is too large the disk stays in a stationary hot state, i.e. a nov\ae-like case. We tried a case where  $\dot{M}_\mathrm{ext}=10^{18}\:\mathrm{g\:s^{-1}}$ and $\beta_\mathrm{eq}=10^3$ and find no eruptions.
 
There may be some other configurations we did not explore where one could find other behaviors. However, since we are not sure of the astrophysical relevance of such a solution we did not investigate this matter further. Indeed, in protoplanetary disks the jet (and so the highly magnetized region of the disk) is coming from the inner region of the disk not the outer region \citep{lee2017}. In DNe, eclipse mapping also reveal an outer disk which is quite in agreement with a typical viscous $\alpha$-disk \citep{horne1985}. 

\end{appendix}

\end{document}